\documentclass[aps,prx,twocolumn,notitlepage,superscriptaddress,showpacs,longbibliography]{revtex4-1}

\usepackage{amsmath}
\usepackage{amsfonts}
\usepackage{amsthm}
\usepackage[utf8]{inputenc}
\usepackage{graphicx}
\usepackage{color}
\usepackage{quantum}
\usepackage[colorlinks]{hyperref}
	\hypersetup{
		bookmarksnumbered,
		pdfstartview={FitH},
		citecolor={blue},
		linkcolor={red},
		urlcolor={black},
		pdfpagemode={UseOutlines}
	}

\newcommand{\bg}[1]{\boldsymbol{#1}}
\newcommand{\inlineheading}[1]{\textit{{#1.---}}}
\newcommand{\mc}[1]{\mathcal{#1}}

\newcommand{\pnat}{\mc{P}_{\mathbb{N}}}
\newcommand{\gpnat}{\mc{GP}_{\mathbb{N}}}
\newcommand{\transp}{\mathrm{T}}

\newtheorem{thm}{Theorem}
\newtheorem{lem}{Lemma}

\begin{document}
\title{Operational resource theory of continuous-variable nonclassicality}

\author{Benjamin Yadin}
\email{benjamin.yadin@gmail.com}
\affiliation{QOLS, Blackett Laboratory, Imperial College London, London SW7 2AZ, United Kingdom}
\affiliation{Clarendon Laboratory, Department of Physics, University of Oxford, Parks Road, Oxford OX1 3PU, United Kingdom}

\author{Felix C. Binder}
\affiliation{
	School of Physical and Mathematical Sciences, Nanyang Technological University, 637371 Singapore, Singapore
}
\affiliation{Complexity institute, Nanyang Technological University, 637335 Singapore, Singapore}

\author{Jayne Thompson}
\affiliation{Centre for Quantum Technologies, National University of Singapore, 3 Science Drive 2, 117543 Singapore, Singapore}

\author{Varun Narasimhachar}
\affiliation{
	School of Physical and Mathematical Sciences, Nanyang Technological University, 637371 Singapore, Singapore
}
\affiliation{Complexity Institute, Nanyang Technological University, 637335 Singapore, Singapore}

\author{Mile Gu}
\email{mgu@quantumcomplexity.org}
\affiliation{
	School of Physical and Mathematical Sciences, Nanyang Technological University, 637371 Singapore, Singapore
}
\affiliation{Complexity institute, Nanyang Technological University, 637335 Singapore, Singapore}
\affiliation{Centre for Quantum Technologies, National University of Singapore, 3 Science Drive 2, 117543 Singapore, Singapore}

\author{M.\ S.\ Kim} \affiliation{QOLS, Blackett Laboratory, Imperial College London, London SW7 2AZ, United Kingdom}

\date{\today}
\begin{abstract}
Genuinely quantum states of a harmonic oscillator may be distinguished from their classical counterparts by the Glauber-Sudarshan P-representation -- a state lacking a positive P-function is said to be nonclassical. In this paper, we propose a general operational framework for studying nonclassicality as a resource in networks of passive linear elements and measurements with feed-forward. Within this setting, we define new measures of nonclassicality based on the quantum fluctuations of quadratures, as well as the quantum Fisher information of quadrature displacements. These lead to fundamental constraints on the manipulation of nonclassicality, especially its concentration into subsystems, that apply to generic multi-mode non-Gaussian states. Special cases of our framework include no-go results in the concentration of squeezing and a complete hierarchy of nonclassicality for single mode Gaussian states.
\end{abstract}

\maketitle

\section{Introduction}
Continuous-variable quantum optical systems exhibit numerous operational advantages over their discrete counterparts, including the unconditional generation of entanglement and relative resilience of such nonclassical states to photon loss.  They have thus played an important role in diverse quantum technologies, spanning computing~\cite{Lloyd1999Quantum,Menicucci2006Universal,Gu2009Quantum,Lund2014Boson}, communication~\cite{Gisin2007Quantum,Ralph1999Continuous,Lance2005Continuous} and metrology
\cite{Paris2009Quantum,Safranek2015Quantum}. Indeed, the reliable creation of entanglement has allowed recent synthesis of ultra-large entangled clusters
that would be difficult to achieve in any other regime~\cite{Yokoyama2013Ultra}.

This divergence in what information processing tasks are considered operationally difficult also motivates different perspectives on nonclassicality. In discrete
variables, nonclassicality is often characterised in terms of coherence and entanglement -- the former to capture the difficulty of creating a quantum
superposition of states in some designated classical basis (such as energy eigenstates), the latter to characterise the difficulty of entangling two
quantum systems. In contrast, in continuous variables, coherent states of light are typically considered the most classical pure states~\cite{Glauber1963Quantum,Sudarshan1963Equivalence} -- owing to their ease of
synthesis. Unlike the energy eigenstates, coherent states are not mutually orthogonal, and represent a superposition of different energies. A state is considered nonclassical when it is not a probabilistic mixture of coherent states~\cite{Mandel1986Non-classical}.
Entanglement itself is considered secondary, as it is easily synthesised by passive linear optics (i.e., networks of beam splitters) once one has a source
of nonclassical light~\cite{Kim2002Entanglement,Wang2002Theorem}. This has motivated tailored means of witnessing and quantifying nonclassicality in the continuous variable
regime~\cite{Vogel2000Nonclassical,Richter2002Nonclassicality,Shchukin2005Nonclassicality,Hillery1987Nonclassical,Lee1991Measure,Marian2002Quantifying,Asboth2005Computable,Gehrke2012Quantification,Vogel2014Unified,Ryl2017Quantifying,Arkhipov2016Nonclassicality}.

A full theoretical understanding of nonclassicality will likely take the form of a resource theory~\cite{Coecke2016Mathematical} -- a mathematical formalism that has enjoyed notable success in the past two decades for describing the structure of entanglement~\cite{Vedral1997Quantifying,Horodecki2009Quantum}. A resource theory gives meaning to the question of \textit{how much} of a quantity of interest is present in a given state. In doing so it renders different states comparable. For instance, in the context of nonclassicality, one might ask if a given squeezed state or a Fock state $\ket{n}$ is more nonclassical -- a question that can be clearly answered with the framework introduced below. Resource theories thus provide a set of criteria for determining whether a proposed quantity counts as a valid measure of the resource. They have been instrumental in understanding quantum reference frames~\cite{Bartlett2007Reference}, thermodynamics~\cite{Horodecki2013Fundamental,Gour2015Resource,Ng2018Resource,Lostaglio2018Thermodynamic}, coherence~\cite{Aberg2006Quantifying,Baumgratz2014Quantifying,Streltsov2017Colloquium}, contextuality~\cite{Grudka2014Quantifying}, steering
\cite{Gallego2015Resource}, and non-Gaussianity~\cite{Genoni2010Quantifying,Takagi2018Convex,Zhuang2018Resource,Albarelli2018Resource}.
Resource-theoretic terminology in continuous variables has appeared in a number of recent works
\cite{Braunstein2005Squeezing,Filip2013Distillation,Rahimi-Keshari2013Quantum,Sabapathy2015Quantum,Sabapathy2016Process,Albarelli2016Nonlinearity,Idel2016Operational,Idel2016Quantum,Tan2017Quantifying},
but these ideas are still in their infancy.

The main contribution of this paper is an operationally motivated resource theory for continuous variable nonclassicality where passive linear optics and measurement
feed-forward are considered operationally simple and thus ``free". We show that this approach naturally leads to a novel quantum resource theory of phase space variance that captures existing views of nonclassicality for both pure and mixed states. For pure states, phase space variance can be
analytically evaluated. For mixed states, we show it can be bounded from below by the quantum Fisher information (QFI) of quadrature displacements. Moreover, we prove that the QFI produces valid indicators of nonclassicality in their own right -- having the same monotone behaviour as phase space variance. This shows a quantitative relation between nonclassicality and performance in metrology. We use this framework to obtain powerful bounds on the concentration of nonclassicality that are applicable to general non-Gaussian multi-mode states. Specialisation of these
techniques to Gaussian states retrieves no-go theorems on the concentration of squeezing, and full hierarchies of nonclassicality in the pure
multi-mode and mixed single-mode cases.

\section{Passive linear optics and the structure of the resource theory}
We work with $n$ bosonic modes with corresponding creation and annihilation operators $a_i^\dagger, a_i$ for $i=1,\dots,n$ satisfying the commutation relations $[a_i,a_j^\dagger]= \delta_{i,j}$. Quadratures operators are defined by $x_i = \frac{1}{\sqrt{2}}(a_i + a_i^\dagger),\, p_i = \frac{1}{\sqrt{2}i} (a_i - a_i^\dagger)$, and may be collected into the vector $\bg q = (q_1,q_2,\dots,q_{2n-1},q_{2n})= (x_1,p_1,\dots,x_n,p_n)$~\cite{Weedbrook2012Gaussian}. The canonical commutators are expressed via $[q_s,q_t] = i \Omega_{st}$, where
\begin{equation}
\Omega = \bigoplus_{i=1}^n \begin{pmatrix}
0 & 1 \\
-1 & 0
\end{pmatrix}
\end{equation}
is the symplectic structure. The quadrature corresponding to a general direction $\bg r \in \mathbb{R}^{2n},\, \abs{\bg r}=1$ in phase space is $\bg r \cdot \bg q$.

In a resource theory viewpoint, one starts from the perspective that, under certain physical conditions or constraints, particular quantum states and processes may be considered resources. The theory is specified by first deciding which sets of states and operations are free, meaning that they have no resource value. A sensible choice of free states is those that are easily prepared in the lab, and similarly free operations are easily performed. This allows for the quantification of resources -- a valid measure $M(\rho)$ of the resource value of a state $\rho$ must satisfy several criteria:
\begin{itemize}
	\item[(i)] $M(\rho) \geq 0$ and vanishes if and only if $\rho$ is a free state;
	\item[(ii)] $M(\rho)$ is nonincreasing when $\rho$ undergoes a free operation;
	\item[(iii)] \emph{convexity}, i.e., $M(\sum_\mu p_\mu \rho_\mu) \leq \sum_\mu p_\mu M(\rho_\mu)$ for any ensemble of states $\rho_\mu$ with probabilities $p_\mu \geq 0,\, \sum_\mu p_\mu =1$.
\end{itemize}	
Property (i) is a natural requirement for ordering resources. Property (ii) expresses the fact that free operations cannot create more of the resource -- this can be expressed in different versions, either for deterministic or probabilistic transformations. Any $M$ satisfying (ii) is called a \emph{monotone}; we reserve the term \emph{measure} for $M$ satisfying both (i) and (ii). Convexity is often desirable since probabilistic mixing is typically considered a free operation.

Here, building on the approach of Ref.~\cite{Tan2017Quantifying}, we choose the free states to coincide with the set $\mc{C}_n$ of classical states on $n$ modes, consisting of convex mixtures of coherent states. In terms of the Glauber-Sudarshan P-function, we have
\begin{equation}
	\mc{C}_n := \left\{ \int \dd^{2n}\bg{\alpha} \; P(\bg{\alpha}) \proj{\bg \alpha} \mathrel{\Big|} P(\bg\alpha) \geq 0 \right\},
\end{equation}
where $\ket{\bg\alpha} := \ket{\alpha_1}\dots \ket{\alpha_n}$ is a product of coherent states, and $\bg\alpha = (\alpha_1,\dots,\alpha_n) \in \mathbb{C}^n$.

The choice of free operations is very important since it determines what it means for one state to be more nonclassical than another. In resource theories generally, it is not always possible to unambiguously say when one state is a more valuable resource than another -- this may depend on the particular task being considered. However, we may say that if $\rho$ can be transformed into $\sigma$ via a free operation, then $\rho$ is at least as nonclassical as $\sigma$.

The set of free unitary operations is easily motivated to be passive linear (PL) unitaries, meaning all number-conserving $U$ such that $[U, \sum_{i=1}^n a_i^\dagger a_i] = 0$, and displacements $D(\bg\alpha) := \prod_{i=1}^n e^{\alpha_i a_i^\dagger - \alpha_i^* a_i}$. These are the most general unitaries mapping the set of classical states $\mc{C}_n$ to itself~\cite{Tan2017Quantifying} -- which is necessary for a consistent description of resources. Moreover, they are operationally ``free" in optics, for instance, where all PL unitaries can be implemented with beam splitters and phase shifters~\cite{Reck1994Experimental}, readily accessible optical elements~\cite{Kok2007Linear}. Displacements are also routinely performed by mixing a state with a large-amplitude coherent state at a beam splitter~\cite{Paris1996Displacement}. By contrast, nonlinear operations are typically very weak and difficult to perform coherently~\cite{Kok2002Single,Andersen2016Thirty}.

The relevant set of general (including nonunitary) free quantum operations is not so clear-cut. In principle, one could take the largest set of classicality-preserving operations (CPOs)~\cite{Gehrke2012Quantification}. However, this is unsatisfactory, lacking a full mathematical characterisation and physical motivation. Instead (following similar arguments for the resource theory of coherence~\cite{Chitambar2016Critical,Yadin2016Quantum,deVicente2017Genuine}) we demand that all nonunitary free operations be implementable with free ancillas and free unitaries, plus measurements and discarding of modes. Therefore we take the free operations to be all possible compositions of the following elements:
\begin{enumerate}
	\item The addition of uncorrelated classical ancilla modes;
	\item PL unitaries and displacements;
	\item destructive measurements on any set of modes;
	\item conditioning on classical randomness, and coarse-graining.
\end{enumerate}
Note that arbitrary positive operator-valued measure (POVM) measurements are permissible as long as they are destructive, i.e., the measured modes are discarded afterwards -- it is easily checked that these preserve classicality. (Note that there is no distinction between entangling and non-entangling measurements, since beam splitters are free.) In most applications, however, only a small subset of measurements may be feasible; for example, when discussing Gaussian states in Section~\ref{sec:gaussian} we limit the discussion to Gaussian measurements. Nevertheless, all our results hold true with more restricted measurements. The final elements in the list simply allow for the conditioning of operations on a classical random number generator and the forgetting of classical information.

As discussed in detail in Appendix~\ref{app:instruments}, our free operations can be described formally as quantum instruments~\cite{Davies1970Operational}. A quantum instrument $\mc{I}$ is a (possibly infinite) collection of completely positive maps $\mc{I} = (\mc{A}_1,\mc{A}_2,\dots)$ such that $\sum_m \mc{A}_m$ is trace-preserving. Each $\mc{A}_m$ describes the operation resulting from a certain selected measurement outcome $m$. Coarse-graining can be described by forming a new instrument $\mc{I'}$ whose elements $\mc{A}'_m$ are sums of distinct partitions of the $\mc{A}_m$. The most fine-grained description of an instrument consists of $\mc{A}_m$ which each have a single associated Kraus operator $K_m$: $\mc{A}_m(\rho) = K_m \rho K_m^\dagger$.

\begin{figure}[h!]
	\centering
	\includegraphics[scale=1]{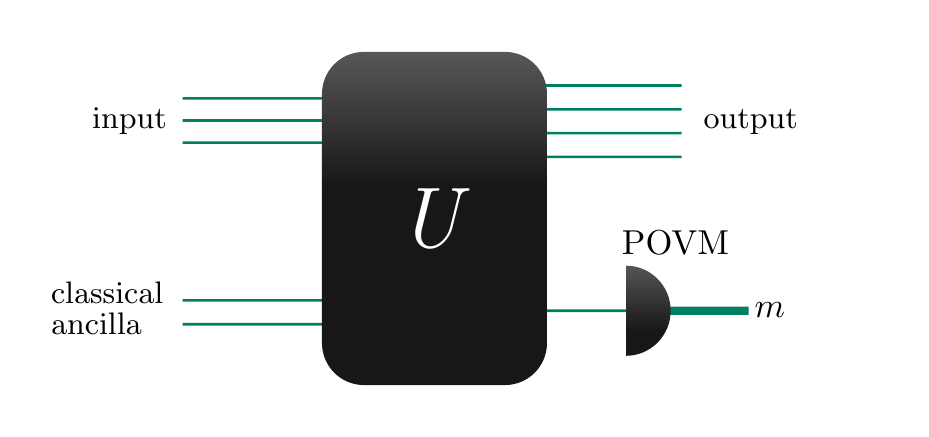}
	\caption{An operation in the set $\mc{P}_1$ constructed with classical ancilla modes, a PL unitary $U$ and a POVM measurement on a set of output modes. $\pnat$ is constructed by repeated concatenation of $\mc{P}_1$ elements with feed-forward depending on each measurement outcome $m$. $\mc{P}_0$ is the special case of $\mc{P}_1$ where the outcome $m$ is not recorded.}
	\label{fig:free_ops}
\end{figure}

We distinguish between different classes of such operations (see Fig.~\ref{fig:free_ops}):
\begin{itemize}
	\item $\mc{P}_0$ is formed by adding a classical ancilla, performing a PL unitary and tracing out a set of modes. An instrument $\mc{I} \in \mc{P}_0$ has a single element.
	
	\item $\mc{P}_1$ is the same as $\mc{P}_0$, but with a measurement first performed on the modes to be traced out.
	
	\item $\mc{P}_r$ is the set of all protocols generated by repeated application of $\mc{P}_1$ with $r$ measurement rounds. Note that feed-forward is permitted.
	
	\item $\pnat := \bigcup_{r=1}^\infty \mc{P}_r$ is the set of all finite-length protocols.
\end{itemize}

Different numbers of input and output modes are permitted in general. In principle, any number of ancilla modes may be used at each stage. Despite this, we prove the following simplification (see Appendix~\ref{app:ancilla_size}):

\begin{lem} \label{lem:ancilla_size}
	The number of ancilla modes for an operation in $\mc{P}_1$ can be assumed to be no larger than the number of output modes.
\end{lem}


Given the chosen sets of free operations, nonclassicality may be viewed generally as a resource in situations where one has access only to the free operations. For example, single-mode nonclassicality is a resource for creating entanglement via passive linear optics: There exists a free operation creating an entangled two-mode state from an input single-mode state $\rho$ if and only if $\rho$ is nonclassical \cite{Kim2002Entanglement,Asboth2005Computable,Killoran2016Converting}. Thus, by examining the structure these operations, we expect to derive general statements about the utility of nonclassical states.


What kind of operations are allowed by taking this free set, rather than the maximal set of CPOs? The following result puts a strong constraint on the action that our free operations can have on coherent states -- it proves that no process in $\pnat$ is able to perform coherent amplification:

\begin{thm}\label{thm:kraus}(No free amplification.)\\
	Every operation in $\pnat$ can be described by a set of Kraus operators $\{K_m\}$ satisfying
	\begin{equation}
		K_m \ket{\bg\alpha} = c_m(\bg\alpha) \ket{M_m \bg\alpha + \bg\delta_m},
	\end{equation}
	where $\bg\alpha \in \mathbb{C}^n$, $c_m(\alpha) \in \mathbb{C}$, $\bg\delta_m \in \mathbb{C}^{n'}$, and $M_m \in \mathbb{C}^{n' \times n}$ has singular values of modulus $\leq 1$. Here $n,\,n'$ are the numbers of input and output modes, respectively.
\end{thm}
(See Appendix~\ref{app:kraus} for full details.) Note that an operation is determined by its action on coherent states thanks to the P-representation. This demonstrates that every free operation can be viewed as a classical mixture of phase space contractions (represented by $M_m$) and displacements (represented by $\bg{\delta}_m$).

Let us consider some examples of CPOs. Single-photon subtraction is a nondeterministic process mapping $\rho \to a \rho a^\dagger$. It is free in our framework, being implementable in $\mc{P}_1$ using a vacuum ancilla, beam splitter and single-photon detection~\cite{Wenger2004Non-gaussian,Parigi2007Probing}. On the other hand, take phase-insensitive noiseless linear amplification, which maps $\ket{\alpha} \mapsto \ket{g \alpha}$ with $\abs{g} > 1$~\cite{Caves1982Quantum}. No deterministic process is able to do this, but it can be done probabilistically with a trace-decreasing map $\mc{E}$ such that $\mc{E}(\proj{\alpha}) = p \proj{g\alpha}$, where $p \leq 1/\abs{g}^2$~\cite{Ralph2009Nondeterministic}. Proposals to implement this utilise either single-photon ancillas (which are nonclassical)~\cite{Ralph2009Nondeterministic} or nonlinear media~\cite{Zavatta2010High}. In fact, Theorem~\ref{thm:kraus} proves that the operation is not in $\pnat$, i.e.,  no free implementation is possible.

Theorem~\ref{thm:kraus} also has implications for the kinds of measures of nonclassicality that are expected in the resource theory. Every monotone under CPOs is necessarily a monotone under the subset $\pnat$ -- but choosing a smaller set of free operations may in principle allow for more monotones. To gain some intuition, consider a cat state $\ket{\psi_c(\alpha)} \propto \ket{\alpha} + \ket{-\alpha}$. Suppose we ask for a process taking this to $\ket{\psi_c(\alpha')}$. Every operation in $\pnat$ is constrained to output $\abs{\alpha'} \leq \abs{\alpha}$, while the above amplification process gives a CPO such that $\abs{\alpha'} > \abs{\alpha}$ with nonzero probability.

Some known measures of nonclassicality will fail to capture this distinction. For example, one can define the distance to the set of classical states, $\inf_{\sigma \in \mc{C}_n} D(\rho,\sigma)$, with some suitable distance measure $D$~\cite{Hillery1987Nonclassical,Marian2002Quantifying}. Taking the trace distance $D(\rho,\sigma) = \frac{1}{2} \tr \abs{\rho-\sigma}$, we obtain a monotone known as the ``nonclassical distance" \cite{Hillery1987Nonclassical}. As shown in \cite{Nair2017Nonclassical}, for $\abs{\alpha} \gg 1$, the dependence on $\alpha$ becomes negligible and saturates at $1/2$, so its monotonicity gives little information about the contractive behaviour. This is essentially due to the near-orthogonality of the branches in that limit. The same is true for the ``entropic entanglement potential", defined as the entropy of entanglement created at a beam splitter, now saturating at $1$ \cite{Asboth2005Computable}. The nonclassical depth defined by Lee~\cite{Lee1991Measure} (namely, the minimal amount of thermal noise that must be added to a state to make it classical), in fact has its maximal value of $1$ for \emph{all} $\alpha$~\cite{Malbouisson2003Measure}. Even more strikingly, the nonclassical distance is found to be decreasing with $\abs{\alpha}$ in the case of an odd cat state $\ket{\alpha}-\ket{-\alpha}$~\cite{Nair2017Nonclassical}. Hence we may ask what kinds of measures truly capture the size of $\abs{\alpha}$ in such superpositions.


\section{Phase space quantum variance}
We propose measures of nonclassicality which capture the separation in phase space of branches of a superposition. In the cat state example, the magnitude of $\alpha$ is an indicator of the macroscopic distinguishability of the branches of the superposition, and hence of the quantum ``macroscopicity" of the state~\cite{Leggett2002Testing,Frowis2017Macroscopic}. This is captured by the maximal variance over all quadratures~\cite{Oudot2015Two,Yadin2016General} -- for a single-mode pure state $\ket{\psi}$, we define
\begin{equation}
	\mc{V}_1(\ket{\psi}) := \max_{\bg r \in \mathbb{R}^2:\; \abs{\bg r}=1} V(\ket{\psi},\bg r \cdot \bg q) - \frac{1}{2},
\end{equation}
where $V(\rho,A) := \tr[\rho A^2] - \tr[\rho A]^2$ is the variance of observable $A$ in the state $\rho$. By definition, $\mc{V}_1$ is invariant under phase rotations. Moreover, it is a faithful witness of nonclassicality, vanishing if and only if $\ket{\psi}$ is classical. This follows from the Heisenberg-Robertson uncertainty principle~\cite{Robertson1929Uncertainty}: For any quadrature $x$ and its conjugate momentum $p$, $V(\ket{\psi},x) V(\ket{\psi},p) \geq 1/4$, so $\mc{V}_1(\ket{\psi}) = 0$ is equivalent to the inequality being saturated with all variances equal to $1/2$; this means $\ket{\psi}$ must be Gaussian~\cite{Stoler1970Equivalence}, and moreover a coherent state.

Alternatively, one can consider the total variance~\cite{Hillery1989Total}, which is related to another measure of macroscopicity~\cite{Lee2011Quantification}:
\begin{equation}
	\mc{W}_1(\ket{\psi}) := V(\ket{\psi},x) + V(\ket{\psi},p) - 1.
\end{equation}
This is again non-negative and is found to be invariant under phase rotations. $\mc{W}_1$ also vanishes if and only if $\ket{\psi}$ is classical. This follows from using the uncertainty relation to write $\mc{W}_1(\ket{\psi}) \geq V(\ket{\psi},x) + 1/[(4V(\ket{\psi},x)] -1$ for any quadrature $x$, with equality if and only if $\ket{\psi}$ is Gaussian. The minimum of the right-hand side is zero, attained at $V(\ket{\psi},x) = 1/2$

For an $n$-mode pure state, we can extend both quantities by considering the covariance matrix, defined for an arbitrary mixed state $\rho$ as
\begin{equation}
	V_{st}(\rho) := \tr \left[\frac{1}{2}\{q_s,q_t\} \rho \right] - \tr[q_s \rho] \tr[q_t \rho],
\end{equation}
In this description, PL unitaries correspond to the set of $2n \times 2n$ orthogonal symplectic matrices $R \in K(n) := O(2n) \cap Sp(2n) \cong U(n)$, namely $R^T R = RR^T = I$ and $R^T \Omega R = \Omega$~\cite{Weedbrook2012Gaussian}. For any phase space direction $\bg r$, we have $V(\rho,\bg r \cdot \bg q) = \bg r^\transp V(\rho) \bg r$.

It follows that $\mc{V}_1$ is identified with the maximal eigenvalue of $V - I/2$. To capture multi-mode structure, we consider the maximal variance over all linear subspaces in phase space of dimension $k \leq 2n$:
\begin{equation}
	\mc{V}_k(\ket{\psi}) := \max_{T:\; \dim T=k} \tr_T [ V(\ket{\psi}) - I/2 ],
\end{equation}
where $\tr_T$ denotes a trace of the matrix restricted to subspace $T$. This is manifestly invariant under free unitaries, and clearly $\mc{V}_1 \leq \mc{V}_2 \leq \dots \leq \mc{V}_{2n}$; all $\mc{V}_k$ vanish exactly on classical states. $\mc{V}_1$ picks out the direction with largest variance, while $\mc{V}_{2n}$ is the total variance over the entire phase space and coincides with the macroscopicity measure of Ref.~\cite{Lee2011Quantification}. $\mc{V}_k$ can be calculated as the sum of the $k$ largest eigenvalues $v_1 \geq \dots \geq v_{2n}$ of $V - I/2$: $\mc{V}_k = \sum_{i=1}^k v_i$~\cite{Horn1985Matrix_ch4}.

For $\mc{W}_k$, we use the concept of a symplectic subspace~\cite{deGosson2006Symplectic}, essentially the $2k$-dimensional subspace of the full phase space $\mathbb{R}^{2n}$ corresponding to a choice of $k \leq n$ modes. Every symplectic subspace can be obtained by applying a rotation $R \in K(n)$ to the subspace spanned by the canonical planes for $(x_1,p_1),\dots,(x_k,p_k)$. We denote the number of modes in a symplectic subspace $\mc{S}$ by $k = \abs{\mc{S}} = \dim \mc{S} /2$. Then
\begin{equation}
	\mc{W}_k(\ket{\psi}) := \max_{\mc{S}:\; \abs{\mc S}=k} \tr_{\mc S}\left[ V(\ket{\psi}) -I/2 \right].
\end{equation}
This is again invariant under free unitaries, $\mc{W}_1 \leq \dots \leq \mc{W}_n$, and each $\mc{W}_k$ vanishes exactly on classical states. Also note that $\mc{W}_k \leq \mc{V}_{2k}$ and $\mc{W}_n = \mc{V}_{2n}$. As shown in Appendix~\ref{app:w_eigenvalues}, we have $\mc{W}_k = \sum_{i=1}^k w_i$, where $w_1 \geq \dots w_n$ are the doubly degenerate eigenvalues of the matrix
\begin{equation}
	W(\ket{\psi}) := \frac{1}{2} \left( V(\ket{\psi}) + \Omega V(\ket{\psi}) \Omega^\transp - I \right).
\end{equation}

\begin{figure}[h!]
	\centering
	\includegraphics[scale=1]{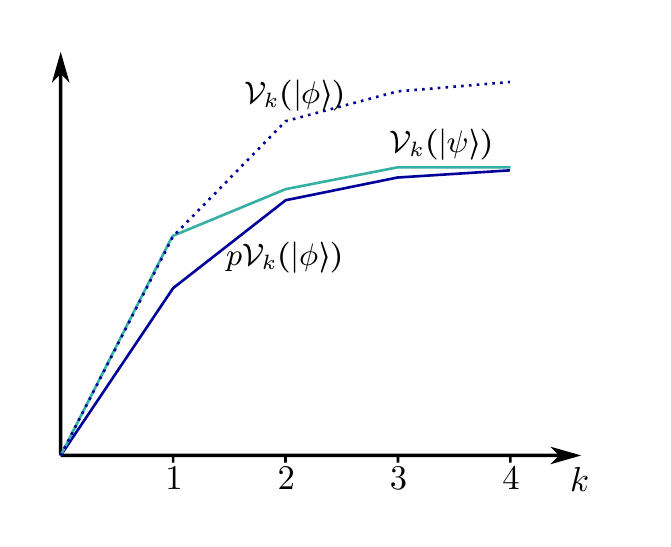}
	\caption{An illustration of the upper bound (\ref{eqn:p1_prob_bound}) on the probability $p$ of reaching a state $\ket{\phi}$ from $\ket{\psi}$ under $\mc{P}_1$. The curve of $\mc{V}_k(\ket{\phi})$ may lie above that of $\mc{V}_k(\ket{\psi})$, but must lie below it when rescaled by $p$.}
	\label{fig:majorisation}
\end{figure}

The covariance matrix is additive under tensor products, $V(\rho \ox \sigma) = V(\rho)\oplus V(\sigma)$, so combining systems amounts to appending one list of $v_i$ to the other and re-ordering. We prove in Appendix~\ref{app:monotones} that the phase space variances $\mc{V}_k,\mc{W}_k$ are monotones for pure state transformations:
\begin{thm}(Monotonicity of phase space variances for pure states.) \label{thm:variance_monotone}
	\begin{itemize}
		\item[(a)] Let $\ket{\psi} \mapsto \ket{\phi}$ with probability p under $\mc{P}_1$, then
		\begin{equation} \label{eqn:vk_monotone}
			p \mc{V}_k(\ket{\phi}) \leq \mc{V}_k(\ket{\psi}) \quad \forall k \leq 2n,
		\end{equation}
		where $n$ is the number of output modes. (If $2k$ is larger than the number of input modes then we take $\mc{V}_k(\ket{\psi}) = \mc{V}_{2n}(\ket{\psi})$.)
		\item[(b)] Let $\ket{\psi} \mapsto \ket{\phi_m}$ with probabilities $p_m$ under $\pnat$, then
		\begin{equation} \label{eqn:vn_monotone}
			\sum_m p_m \mc{V}_{2n}(\ket{\phi_m}) \leq \mc{V}_{2n}(\ket{\psi}).
		\end{equation}
	\end{itemize}
	The same inequalities hold for $\mc{W}_k,\, k \leq n$.
\end{thm}
Note that the total variance $\mc{V}_{2n}$ has been shown to be a stronger monotone than the other $\mc{V}_k$. $\mc{V}_{2n}$ is a full ensemble monotone for arbitrary length adaptive protocols, while feed-forward presents an obstacle to the proof of monotonicity for the remaining quantities. An upper bound on the probability $p$ of a transformation with $\mc{P}_1$ follows from (\ref{eqn:vk_monotone}):
\begin{equation} \label{eqn:p1_prob_bound}
	p \leq \min_{k=1,\dots,2n} \frac{\sum_{i=1}^k v_i(\ket{\psi})}{\sum_{i=1}^k v_i(\ket{\phi})},
\end{equation}
analogous to the condition in entanglement theory derived by Vidal~\cite{Vidal1999Entanglement}. Indeed, (\ref{eqn:vk_monotone}) can be recast as a weak majorization relation: $p \bg{v}(\ket{\phi}) \prec_w \bg{v}(\ket{\psi})$ (see Fig.~\ref{fig:majorisation}). A bound of the same form holds with $v_i$ replaced by $w_i$.

To extend the measures to mixed states, we use a convex roof construction. A state $\rho$ can generally be expressed in many different ways as a classical mixture of pure states: $\rho = \sum_\mu p_\mu \proj{\psi_\mu},\, p_\mu \geq 0,\, \sum_\mu p_\mu = 1$. (We use sum notation for convenience but must bear in mind that a continuous integral may be necessary in general.) The convex roof of $\mc{V}_k$ is the minimal average value of its pure state components, optimised over all possible decompositions of $\rho$:
\begin{equation}
	\hat{\mc V}_k(\rho) := \inf_{\{p_\mu,\ket{\psi_\mu}\}} \sum_\mu p_\mu \mc{V}_k(\ket{\psi_\mu}).
\end{equation}
(We write $\inf$ rather than $\min$ since the noncompactness of the state space makes it unclear whether the infimum is attained.)

$\hat{\mc V}_k$ of course satisfies property (iii), being convex by construction. In addition, this ensures that $\hat{\mc V}_k$ vanishes if $\rho$ is classical. As shown in Appendix~\ref{app:vanishing_condition}, the converse is also true, but the proof is surprisingly nontrivial due to the infimum in the definition. We show this by proving that the trace distance between $\rho$ and $\mc{C}_n$ is upper-bounded by $\hat{\mc V}_1$. Thus $\hat{\mc V}_k$ is a faithful witness of nonclassicality. The same result holds for $\hat{\mc W}_k$.

Furthermore, as shown in Appendix~\ref{app:monotones}, monotonicity carries over from the pure state case, thanks to the convex roof construction. Thus $\hat{\mc V}_k$ and $\hat{\mc W}_k$ are valid measures of nonclassicality for arbitrary states.

One may interpret these measures in terms of quantum fluctuations of quadratures. In the pure state case, they measure the size of quadrature fluctuations above zero-point motion. With mixed states, one must ensure that the quantified fluctuations are quantum in nature and not due to classical uncertainty -- the convexity property (iii) ensures that this is the case.

\section{Metrology monotones}
Unfortunately, $\hat{\mc V}_k$ and $\hat{\mc W}_k$ will in general be hard to calculate, as the lack of a closed form necessitates numerical optimisation. Hence we provide useful lower bounds using the quantum Fisher information. In general, the QFI is defined not for a single state, but for a family of states $\rho_\theta$ parametrised by $\theta \in \mathbb{R}$, and measures the rate of change of $\rho_\theta$ with respect to $\theta$~\cite{Paris2009Quantum}. When the evolution is generated by an observable $H$ via $\rho_\theta = e^{-i\theta H}\rho e^{i\theta H}$, the QFI of a state $\rho$ with respect to $H$ may be defined by $F(\rho,H) := -\partial^2_\theta \left. \mathrm{Fid}(\rho,\rho_\theta) \right|_{\theta=0}$, where $\mathrm{Fid}(\rho,\sigma) = \tr \sqrt{\sqrt{\rho} \sigma \sqrt{\rho}}$ is the fidelity between two states. (For later convenience, our QFI is a factor of four less than the usual definition.)

With multiple observables, a QFI matrix $F$ can be formed. We are interested in the quadratures as generators, and so take matrix elements $F_{st}(\rho) := \frac{1}{4} \tr [\frac{1}{2}\{L_s, L_t\} \rho],\; s,t =1,\dots,2n$, where $L_s$ is the symmetric logarithmic derivative defined implicitly by $\frac{1}{2} \{L_s, \rho\} = -i [q_s, \rho]$. A closed expression can be given in terms of the eigenvectors $\ket{\psi_i}$ and eigenvalues $\lambda_i$ of $\rho$:
\begin{equation} \label{eqn:qfi_formula}
	F_{st} = \frac{1}{2} \sum_{i,j} \frac{(\lambda_i-\lambda_j)^2}{\lambda_i+\lambda_j} \braXket{\psi_i}{q_s}{\psi_j} \braXket{\psi_j}{q_t}{\psi_i} .
\end{equation}
By linearity with respect to the observables, the QFI of any quadrature $\bg{r}\cdot\bg{q}$ is obtained with the quadratic form $F(\rho,\bg{r}\cdot\bg{q}) = \bg{r}^\transp F(\rho) \bg{r}$. The connection with the above measures is the equality for pure states $F(\ket{\psi}) = V(\ket{\psi})$.

For mixed states, the convexity of the QFI~\cite{Toth2013Extremal} again ensures that classical contributions to the variance are not counted. Measures might be constructed analogously to $\mc{V}_k(\ket{\psi})$ by taking traces of $F-I/2$ over linear subspaces. However, the resulting quantities can be negative. To fix this, we take only the positive part by defining
\begin{equation}
	\mc{F}_k(\rho) := \max_{T:\; \dim T=k} \tr_T \left[ F(\rho) - I/2 \right]^+,
\end{equation}
where $[M]^+ = (\abs{M} + M)/2$ denotes the positive part of matrix $M$. In terms of the eigenvalues $f_1 \geq f_2 \dots \geq f_{2n} \geq 0$ of $[F-I/2]^+$, we have $\mc{F}_k = \sum_{i=1}^k f_i$.

Analogously to $\mc{W}_k$, we define
\begin{equation}
	\mc{G}_k(\rho) := \max_{\mc S :\; \abs{\mc S} =k} \tr_{\mc S} \left[ F(\rho) - I/2 \right]^+,
\end{equation}
and $\mc{G}_k = \sum_{i=1}^k g_i$, where $g_1 \geq \dots \geq g_n \geq 0$ are the doubly degenerate eigenvalues of
\begin{equation}
	G(\rho) := \frac{1}{2} \left[ F(\rho) + \Omega F(\rho) \Omega^\transp - I \right]^+.
\end{equation}

These quantities provide useful lower bounds: $\mc{F}_k \leq \hat{\mc V}_k,\, \mc{G}_k \leq \hat{\mc W}_k$. This is because the convex roof is the largest convex function reducing to a specified function on pure states~\cite{Toth2013Extremal}. (Note that, while the QFI with respect to a single direction is the convex roof of the variance~\cite{Yu2013Quantum}, this does not hold for the eigenvalues of the QFI matrix.) It also follows that they vanish for all classical states. The nonclassicality-witnessing property of the QFI has been noted in Ref.~\cite{Rivas2010Precision}. Furthermore, $\mc{F}_k$ is more directly accessible in experiments, there being a number of practical techniques for measuring and lower-bounding QFI~\cite{Frowis2016Detecting,Frowis2016Lower,Girolami2017Witnessing}.

Remarkably, $\mc{F}_k$ and $\mc{G}_k$ are valid monotones, and so put similar constraints on state transformations:

\begin{thm}(Monotonicity of QFI.) \label{thm:qfi_monotone}
	\begin{itemize}
		\item[(a)] Let $\rho \mapsto \sigma$ with probability p under $\mc{P}_1$, then
		\begin{equation} \label{eqn:fk_monotone}
		p \mc{F}_k(\sigma) \leq \mc{F}_k(\rho) \quad \forall k \leq 2n.
		\end{equation}
		Equivalently, $p \bg{f}(\sigma) \prec_w \bg{f}(\rho)$.
		
		\item[(b)] Let $\rho \mapsto \sigma_m$ with probabilities $p_m$ under $\pnat$, then
		\begin{equation} \label{eqn:vn_monotone}
		\sum_m p_m \mc{F}_{2n}(\sigma_m) \leq \mc{F}_{2n}(\rho).
		\end{equation}
		
		\item[(c)] Let $\rho \mapsto \sigma$ under $\mc{P}_0$, then
		\begin{equation}
			f_i(\sigma) \leq f_i(\rho) \quad \forall i \leq 2n.
		\end{equation}
	\end{itemize}
	The same inequalities hold for $\mc{G}_k, g_k,\, k \leq n$.
\end{thm}
Given the operational meaning of the QFI, Theorem~\ref{thm:qfi_monotone} has immediate consequences for metrology: \emph{A performance advantage in sensing quadrature displacements cannot be gained with passive linear optics, measurements and classical resources -- either deterministically or on average in the case of a probabilistic process.} We will see in the next section that this these constraints can be further refined in the context of nonclassicality concentration.

Note also that while $\mc{F}_k$ and $\mc{G}_k$ are monotones, they are not strictly full measures of nonclassicality. This is because they do not satisfy condition (i) -- they can be zero for some non-classical states. An example is given by the state $\rho = (1-p) \sum_{n=1}^\infty p^{n-1} \proj{n}$ (a thermal state with the vacuum term removed). We find $\mc{F}_k(\rho)=0$ when $p > 1/2$ (see Appendix~\ref{app:qfi_nonwitness}), but this state is always nonclassical~\cite{Diosi2000Comment}.

Nevertheless, the monotonicity enables us to still treat these quantities as valid and potentially highly useful quantifiers of nonclassicality. An analogous situation occurs in entanglement theory, where local operations and classical communication (LOCC) operations are considered free. Here, negativity~\cite{Vidal2002Computable,Zychkowski1998Volume} -- a common quantifier of entanglement -- is known to be a monotone under LOCC, but may vanish for certain entangled states~\cite{Zychkowski1998Volume}.

\section{Concentration of nonclassicality and its implications for metrology}

\emph{Concentration of nonclassicality:---} The monotonicity of $\hat{\mc V}_{2n},\hat{\mc W}_n,\mc{F}_{2n},\mc{G}_n$ immediately implies that the total quantum variance cannot increase on average, i.e., the total amount of nonclassicality is not increased by a free operation. However, the monotonicity of $\hat{\mc V}_{k},\hat{\mc W}_k,\mc{F}_{k},\mc{G}_k$ for general $k$ offers a family of more refined constraints. 

Specifically, consider the task of \emph{nonclassicality concentration} --  where a process is to output a subset of modes with higher values of nonclassicality than they originally had, while the remaining modes reduce in value. The above constraints specify that such a task can only be performed with a certain maximal probability.

To illustrate this, consider the following scheme for ``growing cat states"~\cite{Lund2004Conditional}. Starting with a pair of cat states $\ket{\psi_c(\alpha)}^{\ox 2}$ with $\abs{\alpha} \gg 1$, we have $\mc{W}_1 = \abs{\alpha}^2,\, \mc{W}_2 = 2\abs{\alpha}^2$. Interacting the modes at a 50/50 beam splitter gives $\ket{\sqrt{2}\alpha}\ket{0} + \ket{0}\ket{\sqrt{2}\alpha} + \ket{0}\ket{-\sqrt{2}\alpha} + \ket{-\sqrt{2}\alpha}\ket{0}$. Performing a measurement projecting one mode onto the vacuum, with probability $1/2$ the other mode becomes $\ket{\psi_c(\sqrt{2}\alpha)}$, i.e., $\mc{W}_1 = 2\abs{\alpha}^2$ has doubled. While this was observed to be the optimal probability for outputting a double-sized cat state for the scheme in Ref.~\cite{Lund2004Conditional}, Theorem~\ref{thm:variance_monotone}(a) proves that in fact no other protocol can perform better.

Theorem~\ref{thm:qfi_monotone}(c) gives additional, stronger constraints for $\mc{P}_0$ transformations. They highlight that there is a strict hierarchy in terms of the power of the different sets of free operations: concentration of nonclassicality is impossible without measurement. (Theorem~\ref{thm:qfi_gaussian} below extends this to the full $\pnat$ set of operations for the case of Gaussian states.)

\emph{Metrological implications:---}We begin with the operational interpretation of the relevant quantities. First, $\mathcal{F}_1$ represents the optimal performance of a state for sensing a displacement in any single direction (i.e., optimised over all directions in phase space). More generally, $\mathcal{F}_k$ indicates the optimal performance for simultaneous sensing of $k$ orthogonal directions, where an equal weighting is given to each direction. Note that the trace of the QFI is a commonly used figure of merit in multi-parameter metrology~\cite{Szczykulska2016Multi-parameter}. A similar interpretation holds for $\mathcal{G}_k$, where the $2k$ directions are now also required to form a symplectic basis (i.e., $k$ orthogonal conjugate pairs of directions). Theorem~\ref{thm:qfi_monotone}(a) may thus be rephrased as the following statement:

\emph{With only passive linear optics, measurements and classical resources, the conversion of a state which is useful for multi-parameter metrology, into one which is more useful for estimating fewer parameters, necessarily has a limited probability of success.}

Moreover, the eigenvalues $f_i$ of the QFI matrix may be understood as the ability to estimate displacements along the $i$th ``best" direction in phase space for a given state -- i.e., $f_1$ is the most sensitive direction, $f_2$ the second most sensitive, and so on. Let us refer to the corresponding directions in phase space as \emph{principal directions} (note that they form a set of orthogonal axes). Theorem~\ref{thm:qfi_monotone}(c) gives a stronger constraint for  $\mc{P}_0$ transformations than the first parts of the theorem:

\emph{With only passive linear optics and classical resources, the sensitivity to displacements along any principal direction cannot increase.}

\section{Gaussian states} \label{sec:gaussian}
Gaussian states~\cite{Weedbrook2012Gaussian,Adesso2014Continuous,Ferraro2005Gaussian} are fully determined by their first and second quadrature moments. Since displacements are free operations, here we lose no generality by neglecting the first moments $\tr(\rho q_s)$ and characterising states only by their covariance matrices. A simple condition for nonclassicality of Gaussian states is known: $\rho$ is nonclassical if and only if the smallest eigenvalue of $V(\rho)$ satisfies $v_\mathrm{min} < 1/2$~\cite{Simon1994Quantum}. This ``squeezing criterion" simply tests whether there exists a quadrature that exhibits less noise than the vacuum.

It may be shown (see Appendix~\ref{app:gaussian_qfi}) that for Gaussian states, $F = \frac{1}{4} \Omega V^{-1} \Omega^\transp$. Hence all of the above constraints on state transformations can be expressed in terms of eigenvalues of $V$. Moreover, the squeezing criterion implies that $f_1 = 0$ if and only if $\rho$ is classical -- unlike the general case, $f_i,g_i$ and $\mc{F}_k,\mc{G}_k$ are valid measures of nonclassicality for Gaussian states.

We look at the case where input and output both have $n$ modes. Lemma~\ref{lem:ancilla_size} then says that the ancilla used in a $\mc{P}_1$ instrument can be assumed to have $n$ modes (the same conclusion was reached via a different argument in Ref.~\cite{Idel2016Quantum}). While $\mc{P}_0$ necessarily preserves Gaussianity, measurements and feed-forward must be constrained. A Gaussian measurement is described by POVM elements of the form $E(\bg{\alpha}) = \pi^{-n} D(\bg{\alpha}) \Lambda D(\bg{\alpha})^\dagger$, where $\Lambda$ is a Gaussian state~\cite{Adesso2014Continuous}. We constrain the conditional feed-forward operations to be displacements with linear gain (such as those in Ref.~\cite{Zavatta2010High}). We refer to the resulting Gaussian set of free operations as $\gpnat$. This gives a resource theory of squeezing as a sub-resource theory of nonclassicality.

\begin{thm}\label{thm:qfi_gaussian} (No concentration of nonclassicality for Gaussian states.)\\
	Let Gaussian $\rho \mapsto \sigma$ under $\gpnat$ with nonzero probability, then
	\begin{equation}
		f_i(\sigma) \leq f_i(\rho) \quad \forall i \leq 2n
	\end{equation}
	and similarly for $g_i,\, i \leq n$.
\end{thm}
This result shows concentration of nonclassicality is impossible in the Gaussian case (see Appendix~\ref{app:monotones}). It echoes similar no-go results about distillation of resources in Gaussian settings~\cite{Eisert2002Distilling,Fiurasek2002Gaussian,Giedke2002Characterization,Lami2018Gaussian}.

We now determine some necessary and sufficient conditions for Gaussian state transformations. The single-mode case is particularly simple: since $K(1) \cong U(1) \cong SO(2)$, $V$ can be diagonalised with a PL unitary $R \in K(1)$. Hence we only need keep track of its eigenvalues $v_+ \geq v_-$. We define three measures of nonclassicality:
\begin{align}
	\mc{N}_1 & := \max \left\{1 - 2v_-, 0 \right\}, \\
	\mc{N}_2 & := \frac{\mc{N}_1}{2v_+-1}, \\
	\mc{N}_3 & := v_+ \mc{N}_2 = \frac{\mc{N}_1}{2-\frac{1}{v_+}}.
\end{align}
\begin{thm}(Single-mode Gaussian conversion.)
	\begin{itemize}
		\item[(a)] Gaussian $\rho \mapsto \sigma$ under $\mc{P}_0$ if and only if $\mc{N}_1(\sigma) \leq \mc{N}_1(\rho)$ and $\mc{N}_2(\sigma) \leq \mc{N}_2(\rho)$.
		
		\item[(b)] Gaussian $\rho \mapsto \sigma$ under $\gpnat$ with nonzero probability if and only if $\mc{N}_1(\sigma) \leq \mc{N}_1(\rho)$ and $\mc{N}_3(\sigma) \leq \mc{N}_3(\rho)$.
	\end{itemize}
\end{thm}

\begin{figure}[h!]
	\centering
	\includegraphics[scale=1]{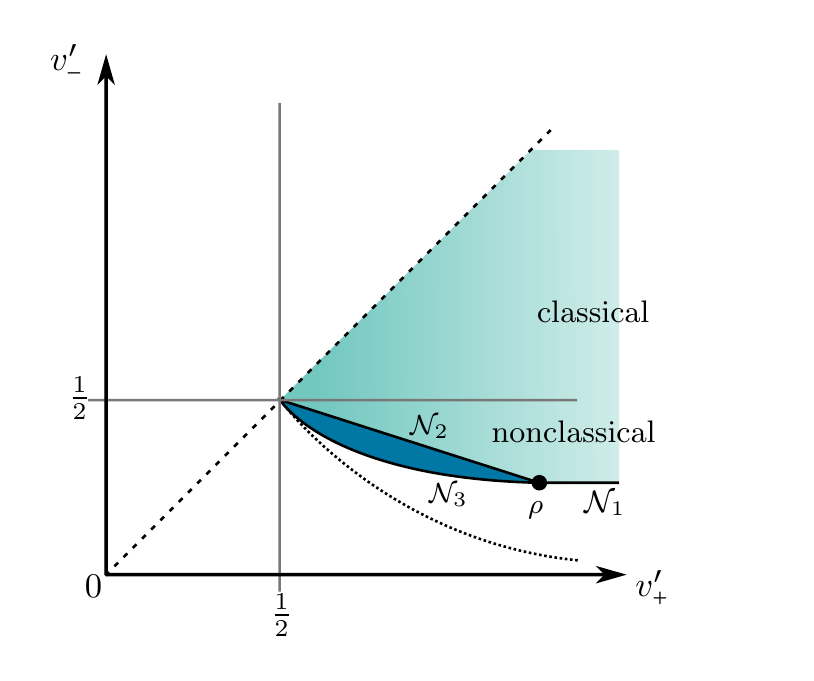}
	\caption{The region of achievable states from a single-mode Gaussian state $\rho$ under free operations. The light shaded region is accessible under $\mc{P}_0$, while $\gpnat$ transformations also access the dark shaded region. The constraints provided by the monotones $\mc{N}_i$ are shown. The dotted curve delimits the physical states satisfying the uncertainty relation $v_+ v_- \geq 1/4$.}
	\label{fig:one_mode_gaussian}
\end{figure}

(See Appendix~\ref{app:gaussian} for the proof.) $\mc{N}_1$ is simply a measure of squeezing, having minimum value zero for classical states and maximum $1$ for infinitely squeezed states. $\mc{N}_2,\, \mc{N}_3$ place limits on the amount of noise that can be removed from the larger-variance quadrature. With $\mc{P}_0$, noise is best removed by mixing with a coherent state at a beam splitter. With $\gpnat$, one can perform a homodyne measurement on the ancilla, thereby reducing the uncertainty in one direction. In both cases, a reduction in noise comes with an associated loss of squeezing.

Under $\mc{P}_0$, a single-mode pure state $\ket{\psi}$ cannot be sent to any pure state other than itself or a coherent state. Since the uncertainty relation is saturated, i.e., $v_+ v_- = 1/4$, we have $\mc{N}_3(\ket{\psi}) = 1/2$ and $\mc{N}_1$ becomes the only nontrivial monotone under $\gpnat$. Thus $\ket{\psi}$ can be transformed under $\gpnat$ into any less squeezed state $\rho$, i.e., with a lower value of $\mc{N}_1$.

For $n$-mode pure states, we again have the simplification that $V$ is diagonalised by $R \in K(n)$:
\begin{equation}
	R^\transp V R = \bigoplus_{i=1}^n \frac{1}{2} \begin{pmatrix}
		s_i & 0 \\ 0 & s_i^{-1}
	\end{pmatrix},
\end{equation}
where the $s_1 \geq \dots \geq s_n \geq 1$ are squeezing parameters~\cite{Giedke2003Entanglement}. Physically, this says that $\ket{\psi}$ can be disentangled into $n$ independent squeezed states $\ket{s_1}\ket{s_2}\dots\ket{s_n}$ by a PL unitary. Thus full conditions for transformations between Gaussian pure states can be given:

\begin{thm} ($n$-mode pure Gaussian conversion.)\\
	A Gaussian pure state $\ket{\psi} \mapsto \ket{\phi}$ under $\gpnat$ if and only if $s_i(\ket{\phi}) \leq s_i(\ket{\psi}) \; \forall i$.
\end{thm}
The necessity of the inequalities follows from Theorem~\ref{thm:qfi_gaussian} after noting that $f_i = s_i/2$ for $i=1,\dots,n$. Conversely, given the inequalities, the above observation on single-mode pure state transformations shows that an operation effecting the transformation exists: diagonalise $\ket{\psi}$ with a PL unitary, operate on each squeezed mode individually, then apply a suitable PL unitary to get $\ket{\phi}$.

\section{Conclusions}
Recent advances in quantum resource theories have led to sophisticated tools for identifying and quantifying nonclassicality -- features of quantum information that distinguish it from classical counterparts. This article has adapted these tools to identify an operational resource theory of nonclassicality in the continuous-variable regime. Our approach was to take those classicality-preserving operations which are considered comparatively easy to engineer in experimental conditions, namely passive linear unitaries and measurements with feed-forward. Quadrature variances and quantum Fisher information emerge naturally within this framework as quantifiers of nonclassicality. These then provide strong and general bounds regarding concentration of quantum resources. In the Gaussian regime, our framework reveals a hierarchy of nonclassicality beyond squeezing. Meanwhile, no-go theorems for squeezing concentration fall out as corollaries, providing a new perspective on recent results in Gaussian resources theories~\cite{Lami2018Gaussian}.

Another noteworthy point is the significance of quantum Fisher information. Commonly used as a measure of performance in parameter estimation, its emergence here indicates an operational interpretation of continuous-variable nonclassicality as a resource for metrological applications. While not every nonclassical state is useful for metrology, we show that ranking one state as more nonclassical than another implies a higher precision in sensing quadrature displacements. Meanwhile, these practical consequences directly lead to experimental means of verifying nonclassicality within a given system, without resorting to full tomography. Quantum Fisher information has also recently been adopted to quantify the macroscopicity of quantum systems~\cite{Frowis2012Measures,Yadin2016General,Frowis2016Lower}. This hints that, in capturing an operational form of nonclassicality for continuous-variable systems, we may naturally recover innate notions of macroscopic quantum effects.

We expect our results to lay the foundations for a systematic understanding of nonclassicality in the continuous-variable regime. Certainly many exciting questions remain. One direction is to extend our result to i.i.d scenarios, where one seeks to convert $N$ copies of a quantum state to another without additional sources of nonclassicality -- in the limit of large $N$. Could we use these ideas to build a hierarchy of nonclassicality for general, possibly non-Gaussian, continuous-variable states? Progress in this direction will ultimately help us fully characterise the distinguishing features of nonclassical light.

\vspace{1em}
\inlineheading{Note added}During preparation of this work, we were made aware of a related paper by Kwon et al.~\cite{Kwon2018Nonclassicality}, which appears concurrently. \\

\section*{Acknowledgements}
We would like to thank Hyukjoon Kwon, Florian Fr\"owis, Howard Wiseman, Syed Assad, Mark Bradshaw, Ping Koy Lam, Tim Ralph and Farid Shahandeh for helpful discussions. This work was supported by UK EPSRC (Grant number  EP/K034480/1 and Doctoral Prize), the KIST Institutional Program (2E26680-18-P025), the National Research Foundation of Singapore (Fellowship NRF-NRFF2016-02), the National Research Foundation and L'Agence Nationale de la Recherche joint project NRF2017-NRF-ANR004 VanQuTe, the Singapore Ministry of Education Tier 1 RG190/17. MSK acknowledges the Samsung GRO project and the Royal Society for their financial support.

%

\bibliography{noncl}

\onecolumngrid

\appendix

\section{Quantum instrument definition of free operations} \label{app:instruments}
Here we give a formal quantum instrument definition of the free operations. We employ the concepts which were introduced in Ref.~\cite{Chitambar2014Everything} to study local operations and classical communication. As mentioned in the main text, a quantum instrument is a family of CP maps $\mc{I} = (\mc{A}_m)_m$ such that $\sum_m \mc{A}_m$ is trace-preserving. Although the inputs and outputs may have different numbers of modes, we will assume that each CP map in a given instrument has the same sized inputs and outputs.

We say that $\mc{I}' = (\mc{A}'_l)_l$ is a coarse-graining of $\mc{I} = (\mc{A}_m)_m$ when the index set of $\mc{I}$ is partitioned into sets $\Sigma_l$ such that $\mc{A}'_l = \sum_{m \in \Sigma_l} \mc{A}_m$.

Any instrument $\mc{I} \in \mc{P}_0$ has a single element: $\mc{I}=(\mc{A})$. Given a set of input system modes $S$ and ancilla modes $A$, we pass the state $\rho_S \ox \rho_A$, where $\rho_A$ is classical, through a PL unitary $U$ and trace out some set of modes $A'$ to give an output $\sigma_{S'}$. Hence
\begin{equation} \label{eqn:p0_dilation}
	\sigma_{S'} = \mc{A}(\rho_S) = \tr_{A'} \left[ U(\rho_S \ox \rho_A) U^\dagger \right].
\end{equation}
For $\mc{P}_1$, we let an arbitrary POVM $\{E^m\}_m,\, E^m \geq 0, \sum_m E^m = I$ act on $A'$. Then
\begin{equation} \label{eqn:p1_dilation}
	\mc{A}_m(\rho_A) = \tr_{A'} \left[ E^m_{A'} U(\rho_S \ox \rho_A) U^\dagger \right].
\end{equation}
To describe multiple rounds with feed-forward, we introduce the following terminology: $\mc{I}' = (\mc{A}'_l)_l$ is PL-linked (passive-linear-linked) to $\mc{I} = (\mc{A}_m)_m$ if there exists a collection of $\mc{J}_m = (\mc{B}_{l|m})_l \in \mc{P}_1$ such that $\mc{I'}$ is a coarse-graining of $(\mc{B}_{l|m} \circ \mc{A}_m)_{m,l}$. Operationally, this means that $\mc{I}$ is performed and the measurement outcome retained; a new $\mc{P}_1$ operation is then performed conditional on the previous measurement; finally, some forgetting of classical information may happen.

Finally, we then say that $\mc{I} \in \mc{P}_r$ for $r \geq 2$ if $\mc{I}$ is PL-linked to an element of $\mc{P}_{r-1}$. The set of all finite-length protocols is $\pnat = \bigcup_{r=1}^{\infty} \mc{P}_r$. Note that, while coarse-graining can be performed after each measurement, it is always possible to assume that this happens only at the end of the protocol. In addition, every protocol can be described as a coarse-graining of a fine-grained instrument in which each $\mc{A}_m$ has a single Kraus operator, $\mc{A}_m(\rho) = K_m \rho K_m^\dagger$.

\section{Maximum required ancilla size} \label{app:ancilla_size}
\begin{figure}[h!]
	\centering
	\includegraphics[scale=1]{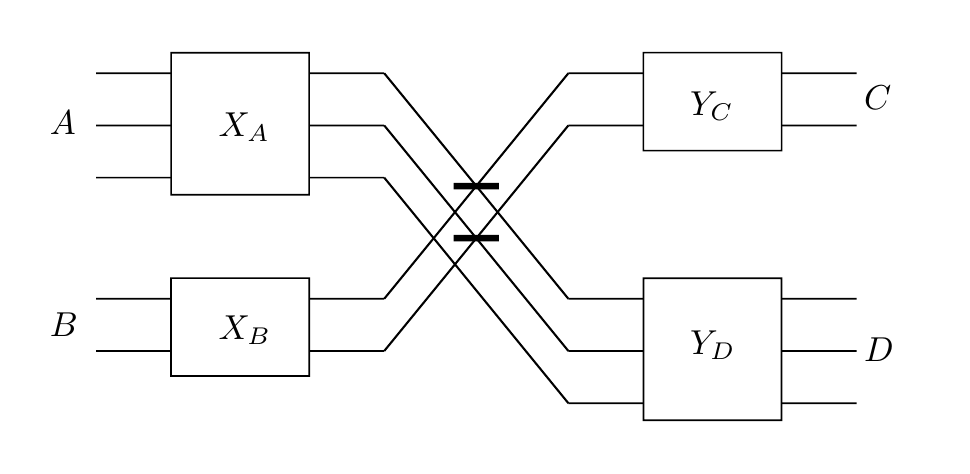}
	\caption{The decomposition of Lemma~\ref{lem:cs_decomp} for the case $(n_A,n_B,n_C,n_D)=(3,2,2,3)$.}
	\label{fig:cs_decomp}
\end{figure}

We first establish a useful statement about the structure of PL unitaries.
\begin{lem} \label{lem:cs_decomp}
	Let $U$ be a PL unitary taking two sets of input modes $A,B$ of sizes $n_A,n_B$ and outputting two sets $C,D$ of sizes $n_C,n_D$. There exists a decomposition of $U$ into $b$ beam splitters taking $A_1 B_1 \to C_1 D_1, \dots, A_b B_b \to C_b D_b$, where $b = \min \{n_A,n_B,n_C,n_D\}$, and initial and final PL unitaries $X_A,X_B, Y_C, Y_D$ acting on each group separately. There remaining modes are transferred between groups. (See Fig.~\ref{fig:cs_decomp}.)
	
	\begin{proof}
		This is a direct consequence of a matrix decomposition result known as the cosine-sine (CS) decomposition~\cite{Paige1994History}. In its most general form, this states that, given partitions $n = n_A + n_B = n_C + n_D$, a matrix $U \in U(n)$ can be written as $U = YDX$, where
		\begin{align}
			X & = \begin{pmatrix}
				X_A  & \\
				& X_B
			\end{pmatrix}, \quad
			Y = \begin{pmatrix}
				Y_C & \\
				& Y_D
			\end{pmatrix}, \\
			D & = \left( \begin{array}{ccc|ccc}
				C	&		&		& S			&		&		\\
				& I		&		&			& O_s^T	&		\\
				&		& O_c	&			&		& I		\\ \hline
				S	&		&		& -C		&		&		\\
				& O_s	&		&			& I		&		\\
				&		& I		&			&		& O_c^T
			\end{array} \right), \\
			C & = \mathrm{diag}(c_1,\dots,c_b), \quad 1 \geq c_1 \geq \dots \geq c_b \geq 0, \\
			S & = \mathrm{diag}(s_1,\dots,s_b), \quad 0 \leq s_1 \leq \dots \leq s_b \leq 1, \\
			& C^2 + S^2 = I.
		\end{align}
		Here, $X_A,X_B,Y_C,Y_D$ are unitaries of respective dimensions $n_A,n_B,n_C,n_D$ and $D$ is shown partitioned into blocks of $n_C,n_D$ rows and $n_A,n_B$ columns. The identity matrices $I$ are not all the same size, and $O_c,O_s$ are rectangular matrices of zeroes whose sizes depend on the partition.

		$X$ and $Y$ correspond to the initial and final ``local" unitaries. To interpret the form of $D$, first note that the sub-block containing the $C$ and $S$ matrices corresponds to beam splitters of reflectivities $c_i^2$ between the first $b$ modes of $A$ and $B$. The remaining sub-blocks correspond to direct transferral of modes from $A \to C$ and $B \to D$, as well as swapping $A \to D$ and $B \to C$.
	\end{proof}
\end{lem}

Now consider a $\mc{P}_1$ operation taking input system $S$ and ancilla $A$ to output system $S'$ and ancilla $A'$, where a final measurement is performed on $A'$. We first show that, if $n_{A} > n_{S'}$, then there are ancilla modes which do not interact with the system. Later, we will show that these are unnecessary. (In the following, we can ignore the local PL unitaries $X_S,X_A,Y_{S'},Y_{A'}$ acting on each group without loss of generality.) Suppose first that $n_{S} \geq n_{S'}$. Then, by applying Lemma~\ref{lem:cs_decomp} to the input groupings $S,A$ and output groupings $S',A'$, at most $n_{S'}$ beam splitters are needed. So only $n_{S'}$ modes in $A$ are required for the beam splitters, while the remainder go directly to $A'$. If instead $n_{S} \leq n_{S'}$, then only $n_{S}$ beam splitters are needed. $n_{S}$ of the outputs from the beam splitters go to $S'$; to make up the remainder, $n_{S'} - n_{S}$ modes from $A$ must be transferred to $S'$. So in total, no more than $(n_{S'}-n_S) + n_S = n_{S'}$ modes are needed for $A$, apart from those going directly into $A'$.

\begin{figure}[h!]
	\centering
	\includegraphics[scale=1]{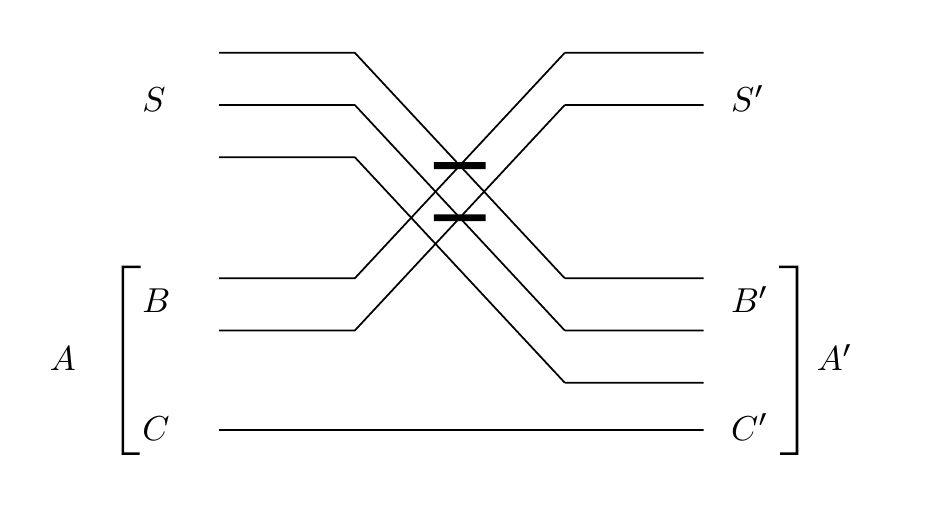}
	\caption{An example of unnecessary ancilla modes ($C,C'$) for a $\mc{P}_1$ operation.}
	\label{fig:ancilla_modes}
\end{figure}

Suppose that $n_A > n_{S'}$. By the above argument, we can divide $A$ into two sets of modes $B,C$ and similarly $A'$ into $B',C'$, such that $B$ interacts with $S$ via beam splitters with possible transferral of modes, $C$ maps directly onto $C'$, and $B'$ contains all other measured modes (see Fig.~\ref{fig:ancilla_modes}). Since the ancilla is initially classical, we can write
\begin{equation}
	\rho_{BC} = \int \dd^{2n_{C'}} \bg\gamma \; P(\bg \gamma) \rho_{B|\bg\gamma} \ox {\proj{\bg\gamma}}_C, \quad P(\bg\gamma) \geq 0,
\end{equation}
where $\ket{\bg \gamma}$ is a coherent state on $C$. Let $\sigma_{S'B'|\bg\gamma}$ be the result of applying the PL unitary to $\rho_S \ox \rho_{B|\bg\gamma}$ and transferring any necessary modes from $S$ to $B'$ and $B$ to $S'$. This describes the dynamics of all modes apart from $C$ going to $C'$. Thus, with a POVM $\{E^m_{B'C'}\}$ , the output state is
\begin{align}
	p_m \sigma_{S'|m} & = \tr_{B'C'} \left[ E^m_{B'C'} \int \dd^{2n_{C'}} \bg\gamma \; P(\bg\gamma) \sigma_{S'B'|\bg\gamma} \ox {\proj{\bg\gamma}}_{C'}  \right] \\
	& = \tr_{B'} \left[ \int \dd^{2n_{C'}} \bg\gamma \; P(\bg\gamma) F^m_{B'|\bg\gamma} \sigma_{S'B'|\bg\gamma} \right],
\end{align}
where $F^m_{B'|\bg\gamma} := {\bra{\bg\gamma}}_{C'} E^m_{B'C'} {\ket{\bg\gamma}}_{C'}$ defines a new POVM for each $\bg\gamma$. Thus we see that the $C$ modes are unnecessary and just result in a classical mixture of different $\mc{P}_1$ protocols.

\section{Kraus operators for $\pnat$} \label{app:kraus}
First note that every quantum operation is uniquely determined by its action on coherent states, due to the P-representation and the linearity of operations. So we only need to consider fine-grained elements of $\pnat$, in which all ancilla states are initially pure (i.e., coherent states) and all measurements are rank-1 POVMs. We look at a single such element of $\mc{P}_1$ with input ${\ket{\bg\alpha}}_S$ and initial ancilla state ${\ket{\bg\beta}}_A$. Before the measurement, we have the state ${\ket{\bg{\alpha'}}}_{S'} {\ket{\bg{\beta'}}}_{A'}$, where
\begin{equation}
	\begin{pmatrix}
		\bg{\alpha'} \\ \bg{\beta'}
	\end{pmatrix} =
	\begin{pmatrix}
		U_{11} & U_{12} \\
		U_{21} & U_{22}
	\end{pmatrix}
	\begin{pmatrix}
		\bg\alpha \\ \bg\beta
	\end{pmatrix},
\end{equation}
where $U_{11}$ is $n_{S'} \times n_{S}$, $U_{12}$ is $n_{S'} \times n_{A}$, and so on. It is immediately clear from the decomposition in Lemma~\ref{lem:cs_decomp} that $U_{11}$ has singular values of modulus $\leq 1$. Hence $\bg{\alpha'} = U_{11} \bg\alpha + U_{12} \bg\beta$, where $U_{11}$ is a contraction, and the displacement by $U_{12}\bg\beta$ is independent of the input $\bg\alpha$.

Since ${\ket{\bg{\alpha'}}}_{S'} {\ket{\bg{\beta'}}}_{A'}$ is uncorrelated, any measurement on $A'$ has no back-action on $S'$. Suppose a rank-1 POVM has outcome $m$ projecting $A'$ onto $\ket{\psi_m}$, then the Kraus operator determining the mapping from input to output is
\begin{equation}
	K_m {\ket{\bg\alpha}}_{S} = c_m(\bg\alpha) {\ket{U_{11}\bg\alpha + U_{12}\bg\beta}}_{S'},
\end{equation}
where $c_m(\bg\alpha) = \braket{\psi_m}{U_{21}\bg\alpha + U_{22}\bg\beta}$.

Finally, it is clear that the Kraus operator form remains the same under multiple rounds with feed-forward, where the contractions and displacements applied may depend on previous measurement outcomes. Hence the stated form of general $K_m$ for $\pnat$.

\section{Eigenvalue expression for $\mc{W}_k$} \label{app:w_eigenvalues}
$W$ has a particular structure in terms of $2 \times 2$ blocks:
\begin{equation}
	W = \begin{pmatrix}
		W^{11} & W^{12} & \dots \\
		W^{21} & W^{22} & \dots \\
		\vdots & \vdots & \ddots
		\end{pmatrix}, \quad
		W^{ij} = \begin{pmatrix}
		W^{ij}_R & -W^{ij}_I \\
		W^{ij}_I & W^{ij}_R
	\end{pmatrix},
\end{equation}
where each diagonal block has $W^{ii}_I = 0$. This lets us use an isomorphism~\cite{deGosson2012Symplectic} onto a complex vector space of half the dimension: We form $\tilde W \in \mathbb{C}^{n\times n}$ with elements $\tilde{W}_{ij} := W^{ij}_R + i W^{ij}_I$, and similarly a vector $\bg r = (r_{1,x},r_{1,p},r_{2,x},r_{2,p},\dots) \in \mathbb{R}^{2n}$ is mapped to $\bg{\tilde r} = (r_{1,x} + i r_{1,p}, r_{2,x} + i r_{2,p},\dots) \in \mathbb{C}^n$. Then $\bg{\tilde r}^\dagger \tilde{W} \bg{\tilde r} = \bg{r}^\transp W \bg{r}$; in addition, an orthogonal basis in $\mathbb{C}^n$ corresponds to a symplectic basis in $\mathbb{R}^{2n}$. Therefore
\begin{equation}
	\mc{W}_k = \max_{T:\, \dim T = k} \tr_{T} \tilde W,
\end{equation}
where the maximisation is over all subspaces $T$ of complex dimension $k$.

Note that $\tilde W$ is hermitian and positive semi-definite; the Courant-Fischer theorem~\cite{Horn1985Matrix_ch4} states that its eigenvalues $w_1 \geq w_2 \geq \dots w_n \geq 0$ can be obtained by
\begin{align}
w_i &= \max_{T:\, \dim T=i} \, \min_{\bg{c} \in T :\, \abs{\bg c} = 1} \bg{c}^\dagger \tilde{W} \bg{c} \\
&= \max_{\mc{S}:\, \abs{\mc S}=i} \, \min_{\bg{r} \in \mc{S} :\, \abs{\bg r} = 1} \bg{r}^\transp W \bg{r},
\end{align}
and moreover $\mc{W}_k = \sum_{i=1}^k w_i$. The fact that these are the doubly degenerate eigenvalues of $W$ is evident from inverting the isomorphism to map from the diagonal form of $\tilde W$ back to the real $2n$-dimensional matrix $\mathrm{diag}(w_1,w_1,w_2,w_2,\dots)$.

\section{Vanishing of convex roof} \label{app:vanishing_condition}
Here we show that $\hat{\mc V}_k(\rho) = 0 \Rightarrow \rho \in \mc{C}_n$. This is not straightforward because of the infimum in the definition. We start with an inequality showing that small $\hat{\mc V}_k(\rho)$ implies $\rho$ is close to $\mc{C}_n$. We use the nonclassical distance as defined in Ref.~\cite{Hillery1987Nonclassical}:
\begin{equation}
	\delta(\rho) := \inf_{\sigma \in \mc{C}_n} D_{\tr}(\rho,\sigma),
\end{equation}
where $D_{\tr}(\rho,\sigma) := \frac{1}{2} \tr \abs{\rho-\sigma}$ is the trace distance~\cite{Nielsen2010Quantum_ch9}.
\begin{lem} \label{lem:distance_variance_bound}
	For an $n$-mode pure state $\ket{\psi}$, there exists a classical pure state $\ket{\phi}$ such that
	\begin{equation}
		D_{\tr}(\ket{\psi},\ket{\phi})^2 \leq n \mc{V}_1(\ket{\psi}).
	\end{equation}
	\begin{proof}
		We initially assume for simplicity that $\ket{\psi}$ has vanishing first moments. Choose a set of modes with quadratures $x_i,p_i$ such that $x_1$ is the optimal quadrature with $\mc{V}_1(\ket{\psi}) = V(\ket{\psi},x_1)-1/2$. Defining number operators $N_i = (x_i^2 + p_i^2 - 1)/2$ and $N = \sum_{i=1}^n N_i$, we have
		\begin{align}
			\expect{N} & = \frac{1}{2} \sum_{i=1}^n \left[ V(\ket{\psi},x_i) + V(\ket{\psi},p_i) - 1 \right] \\
				& \leq n \left( V(\ket{\psi},x_1) - \frac{1}{2} \right) \\
				& = n \mc{V}_1(\ket{\psi}).
		\end{align}
		The overlap between $\ket{\psi}$ and the $n$-mode vacuum state is $\abs{\braket{0}{\psi}}^2 = P(N=0)$. By writing $\expect{N} = \sum_{k=0}^{\infty} k P(N=k) \geq \sum_{k=1}^\infty P(N=k)$, we obtain
		\begin{align}
			\abs{\braket{0}{\psi}}^2 = 1 - P(N \geq 1) \geq 1 - \expect{N} \geq 1 - n\mc{V}_1(\ket{\psi}).
		\end{align}
		For pure states, the trace distance simplifies to $D_{\tr}(\ket{\psi},\ket{0}) = \sqrt{1-\abs{\braket{0}{\psi}}^2}$~\cite{Nielsen2010Quantum_ch9}, hence $D_{\tr}(\ket{\psi},\ket{0})^2 \leq n\mc{V}_1(\ket{\psi})$. Finally, in general we take $\ket{\phi}$ to be the coherent state with the same first moments as $\ket{\psi}$.
	\end{proof}
\end{lem}

If $\hat{\mc V}_1(\rho) = 0$, then there exists a sequence of pure state decompositions such that $\rho = \sum_\mu p_{\alpha,\mu} \proj{\psi_{\alpha,\mu}}$ for each $\alpha = 0,1,2,\dots$, and $\lim_{\alpha \to \infty} \sum_\mu p_{\alpha,\mu}\mc{V}_1(\ket{\psi_{\alpha,\mu}}) = 0$. So for any $\epsilon > 0$, there exists $\alpha^*$ such that
\begin{equation}
	\sum_\mu p_{\alpha,\mu} \mc{V}_1(\ket{\psi_{\alpha}}) \leq \epsilon \quad \forall \alpha \geq \alpha^*.
\end{equation}
Applying Lemma~\ref{lem:distance_variance_bound}, we find a sequence of classical $\ket{\phi_{\alpha,\mu}}$ such that
\begin{equation}
	\sum_\mu p_{\alpha,\mu} D_{\tr}(\ket{\psi_{\alpha,\mu}},\ket{\phi_{\alpha,\mu}})^2 \leq n \epsilon \quad \forall \alpha \geq \alpha^*.
\end{equation}
The Cauchy-Schwarz inequality gives
\begin{align}
	\left( \sum_\mu p_{\alpha,\mu} D_{\tr}(\ket{\psi_{\alpha,\mu}},\ket{\phi_{\alpha,\mu}}) \right)^2 & = \left( \sum_\mu \sqrt{p_{\alpha,\mu}} \times \sqrt{p_{\alpha,\mu}} D_{\tr}(\ket{\psi_{\alpha,\mu}},\ket{\phi_{\alpha,\mu}}) \right)^2 \\
		& \leq \left( \sum_\mu p_{\alpha,\mu} \right) \left( \sum_\mu p_{\alpha,\mu} D_{\tr}(\ket{\psi_{\alpha,\mu}},\ket{\phi_{\alpha,\mu}})^2 \right) \\
		& = \sum_\mu p_{\alpha,\mu} D_{\tr}(\ket{\psi_{\alpha,\mu}},\ket{\phi_{\alpha,\mu}})^2.
\end{align}
The convexity of the trace distance~\cite{Nielsen2010Quantum_ch9} then gives
\begin{equation}
	D_{\tr}(\rho, \sigma_\alpha)^2 \leq \left( \sum_\mu p_{\alpha,\mu} D_{\tr}(\ket{\psi_{\alpha,\mu}},\ket{\phi_{\alpha,\mu}}) \right)^2 \leq n \epsilon,
\end{equation}
where $\sigma_\alpha := \sum_\mu p_{\alpha,\mu} \proj{\phi_{\alpha,\mu}}$. Therefore, by choosing sufficiently large $\alpha$, we can find a classical state $\sigma_\alpha$ that is arbitrarily close to $\rho$ in trace distance. Hence $\delta(\rho) = 0$. As shown in Ref.~\cite{Nair2017Nonclassical}, this implies that $\rho$ is classical.

Finally, we note that the same conclusions hold for $\hat{\mc W}_k$. Instead of the inequality $\expect{N} \leq n \mc{V}_1$, we use $\expect{N} \leq n \mc{W}_1$.

It is also worth noting that states can be close to classical in trace distance but have arbitrarily large variance. For example, take a superposition of number states $\ket{\psi_l} = \sqrt{1-\epsilon} \ket{0} + \sqrt{\epsilon} \ket{l}$ with $l>2$. Now $\abs{\braket{0}{\psi_l}}^2 = 1-\epsilon$ but $V(\ket{\psi_l},x) = 1/2 + \epsilon l$, which can be made arbitrarily large by choosing large enough $l$.

\section{Convexity of QFI measures}
The QFI of any single observable $A$ is convex: for any ensemble of states $\rho_\mu$ with probabilities $p_\mu$, $F(\sum_\mu p_\mu \rho_\mu,A) \leq \sum_\mu p_\mu F(\rho_\mu,A)$. Then
\begin{align}
	\mc{F}_k(\sum_\mu p_\mu \rho_\mu) & \leq \max_{T:\; \dim T=k} \sum_\mu p_\mu \tr_T \left[F(\rho_\mu)-I/2 \right]^+ \\
		& \leq \max_{\{T_\mu:\; \dim T_\mu=k\}} \sum_\mu p_\mu  \tr_T \left[F(\rho_\mu)-I/2 \right]^+ \\
		& = \sum_\mu p_\mu \mc{F}_k(\rho_\mu).
\end{align}
The same applies to $\mc{G}_k$.

\section{Nonclassical state not witnessed by QFI} \label{app:qfi_nonwitness}
We give an example of a state whose nonclassicality is not witnessed by the QFI measure: $\rho = (1-p) \sum_{n=1}^\infty p^{n-1} \proj{n}$. This was used in a proof~\cite{Diosi2000Comment} of lack of sufficiency of a nonclassicality witness by Vogel~\cite{Vogel2000Nonclassical}, and is nonclassical for all $p \in (0,1)$. Using the expression (\ref{eqn:qfi_formula}), we have
\begin{align}
	F(\rho,x) & = \left( \frac{1-p}{p} \right) \sum_{n>m\geq 1}^\infty \frac{(p^n-p^m)^2}{p^n+p^m} \abs{\braXket{n}{x}{m}}^2 + \left( \frac{1-p}{p} \right) \sum_{m \geq 1}^\infty p^m \abs{\braXket{0}{x}{m}}^2 \\
		& = \left( \frac{1-p}{p} \right) \left[ \sum_{m=1}^\infty \frac{(p^{m+1}-p^m)^2}{p^{m+1}+p^m} \frac{(m+1)}{2} + \frac{p}{2} \right] \\
		& = \left( \frac{1-p}{2p} \right) \left[ p + \sum_{m=1}^\infty p^m \frac{(1-p)^2}{1+p} (m+1) \right] \\
		& = \frac{1-p}{2} + \frac{(1-p)^3}{2p(1+p)} \left[ \sum_{n=0}^\infty n p^{n-1} - 1 \right].
\end{align}
(Note that this state is symmetric with respect to phase rotations, so $x$ can be any quadrature.) After some algebra, we get
\begin{equation}
	F(\rho,x) = \frac{3}{2} \left( \frac{1-p}{1+p} \right),
\end{equation}
and $F(\rho,x) < 1/2$ for $p > 1/2$ -- so the QFI does not detect the nonclassicality of this state.

\section{Monotonicity proofs} \label{app:monotones}
We start by proving monotonicity of $\mc{V}_k$ for pure states by checking the behaviour under each of the elementary operations.
\begin{enumerate}
	\item Addition of uncorrelated classical ancilla modes:\\
	The covariance matrix for a product of states is simply the direct sum $V({\ket{\psi}}_S {\ket{\bg \alpha}}_A) = V({\ket{\psi}}_S) \oplus V({\ket{\bg \alpha}}_A)$ with $V({\ket{\bg \alpha}}_A) = I/2$. Hence the vector $\bg v({\ket{\psi}}_S {\ket{\bg \alpha}}_A)$ is just $\bg v({\ket{\psi}}_S)$ with zeros appended. So $\mc{V}_k$ is unchanged.
	
	\item PL unitaries and displacements:\\
	The eigenvalues of $V$ are manifestly invariant under these unitaries.
	
	\item Destructive measurements:\\
	We divide the whole set of modes into $S$ and $A$, where $A$ is to be measured; let $T_S, T_A$ be their respective (orthogonal) subspaces of the total phase space, such that $T_S \oplus T_A = \mathbb{R}^{2(n_S+n_A)}$. Taking any global state ${\ket{\psi}}_{SA}$ and $k \leq 2n_S$,
	\begin{equation}
		\mc{V}_k({\ket{\psi}}_{SA}) = \max_{T:\; \dim T=k} \tr_T \left[V({\ket{\psi}}_{SA}) - I/2 \right]^+ \geq \max_{T \subseteq T_S:\; \dim T=k} \tr_T \left[V({\ket{\psi}}_{SA}) - I/2 \right]^+.
	\end{equation}
	Now we use a crucial property of the variance~\cite{Schuch2004Quantum}: for any observable $x_S$ acting only on $S$ and (rank-1) POVM acting only on $A$,
	\begin{equation} \label{eqn:var_monotone}
		V({\ket{\psi}}_{SA}, x_S) \geq \sum_m p_m V({\ket{\phi_m}}_S, x_S),
	\end{equation}
	where the ${\ket{\phi_m}}_S$ is the state resulting from measurement outcome $m$ with probability $p_m$. This extends to a trace of the covariance matrix over any subspace $T \subseteq T_S$, hence
	\begin{equation} \label{eqn:vk_intermediate}
		\mc{V}_k({\ket{\psi}}_{SA}) \geq \max_{T \subseteq T_S :\; \dim T=k} \sum_m p_m \tr_T \left[ V({\ket{\phi_m}}_S) - I/2 \right]^+.
	\end{equation}
	For any single $m$, we therefore have
	\begin{equation}
		\mc{V}_k({\ket{\psi}}_{SA}) \geq p_m \max_{T \subseteq T_S :\; \dim T = k} \tr_T \left[ V({\ket{\phi_m}}_S) - I/2 \right]^+ = p_m \mc{V}_k({\ket{\phi_m}}_S).
	\end{equation}
	In the case $k = 2n_S$, there is no maximisation on the right-hand side of (\ref{eqn:vk_intermediate}), so
	\begin{equation}
		\mc{V}_{2n_S}({\ket{\psi}}_{SA}) \geq \sum_m p_m \tr_{T_S} \left[ V({\ket{\phi_m}}_S) - I/2 \right]^+ = \sum_m p_m \mc{V}_{2n_S}({\ket{\phi_m}}_S).
	\end{equation}
	
	\item Classical randomness and coarse-graining:\\
	In the pure state case, the only allowed such operation is to coarse-grain measurement outcomes that give the same output (otherwise mixed states are produced) -- this changes nothing.
\end{enumerate}

It is easily checked that all of the above logic works identically for $\mc{W}_k$, the only difference being that maximisation is now performed over symplectic subspaces of dimension $2k$.

For the convex roof $\hat{\mc V}_k$, elements (1) and (2) work as above, so we first address (3). For any $\epsilon > 0$, we can find a pure state decomposition $\rho_{SA} = \sum_\mu r_\mu {\proj{\psi_\mu}}_{SA}$ such that $\sum_\mu r_\mu \mc{V}_k( {\ket{\psi_\mu}}_{SA}) \leq \hat{\mc V}_k(\rho_{SA}) + \epsilon$. Let the POVM on $A$ act on ${\ket{\psi_\mu}}_{SA}$ to give ${\ket{\phi_{m|\mu}}}_S$ with probability $p_{m|\mu}$. For any fixed $m$,
\begin{align}
	\hat{\mc V}_k(\rho_{SA}) + \epsilon & \geq \sum_\mu r_\mu p_{m|\mu} \mc{V}_k({\ket{\phi_{m|\mu}}}_S) \\
		& \geq  p_m \hat{\mc V}_k(\sigma_{S|m}),
\end{align}
where $p_m \sigma_{S|m} = \sum_\mu r_\mu p_{m|\mu} {\proj{\phi_{m|\mu}}}_S$ is the state obtained from measurement of $\rho_{SA}$ with probability $p_m$. Letting $\epsilon \to 0$ gives the desired result. In the case $k = 2n_S$, we have the stronger inequality $\hat{\mc V}_{2n_S}(\rho_{SA}) \geq \sum_m p_m \hat{\mc V}_{2n_S}({\rho_m}_S)$. \\

Monotonicity under coarse-graining simply follows from convexity of $\hat{\mc V}_k$. \\

Again, everything works analogously for $\hat{\mc W}_k$. \\

We now prove monotonicity for $\mc{F}_k$. The QFI matrix of a product state is again a direct sum $F(\rho_S \ox \rho_A) = F(\rho_S) \oplus F(\rho_A)$ due to additivity of the QFI~\cite{Toth2014Quantum}, with $F(\rho_A) \leq I/2$ if $\rho_A$ is classical. Invariance under PL unitaries is due to the quadratic form $F(\rho,\bg r \cdot \bg q) = \bg r^\transp F(\rho) \bg r$. Displacements have the action $q \to q + \text{constant}$ for any quadrature $q$, under which $F(\rho,q)$ is unchanged.

Part (3) hinges on the property of the QFI analogous to (\ref{eqn:var_monotone})~\cite{Yadin2016General}:
\begin{equation} \label{eqn:qfi_monotone}
	F(\rho_{SA}, x_S) \geq \sum_m p_m F(\sigma_{S|m}, x_S).
\end{equation}
Then the proofs of monotonicity of $\mc{F}_k$ and $\mc{G}_k$ proceed exactly as above. \\

For special cases where the QFI matrix is the same for every outcome of a measurement, a stronger constraint can be given: the monotonicity of the eigenvalues of the QFI matrix. This follows from the Courant-Fischer theorem~\cite{Horn1985Matrix_ch4},
\begin{equation}
	f_i =  \max_{T:\, \dim T =i} \, \min_{\bg{r} \in T :\, \abs{\bg r} = 1} \bg{r}^\transp \left[F - I/2 \right]^+ \bg{r}.
\end{equation}
Following the same logic as above, for any initial state $\rho_{SA}$ and a measurement on $A$, we have
\begin{align}
	f_i(\rho_{SA}) & = \max_{T:\, \dim T =i} \, \min_{\bg{r} \in T :\, \abs{\bg r} = 1} \left[ F(\rho_{SA}, \bg{r}\cdot\bg{q})-1/2 \right]^+ \\
		& \geq \max_{T \subseteq T_S:\, \dim T =i} \, \min_{\bg{r} \in T :\, \abs{\bg r} = 1} \left[ F(\rho_{SA}, \bg{r}\cdot\bg{q})-1/2 \right]^+ \\
		& \geq \max_{T \subseteq T_S :\, \dim T =i} \, \min_{\bg{r} \in T :\, \abs{\bg r} = 1} \sum_m p_m \left[ F(\sigma_{S|m}, \bg{r}\cdot\bg{q})-1/2 \right]^+.
\end{align}
When all outcomes have the same QFI matrix independent of $m$, for any $m$ we have
\begin{equation}
	f_i(\rho_{SA}) \geq \max_{T \subseteq T_S :\, \dim T =i} \, \min_{\bg{r} \in T :\, \abs{\bg r} = 1} \left[ F(\sigma_{S|m}, \bg{r}\cdot\bg{q})-1/2 \right]^+ = f_i(\sigma_{S|m}).
\end{equation}
This applies both to $\mc{P}_0$ (since the measurement is trivial) and to Gaussian states with Gaussian measurements (see Appendix~\ref{app:gaussian}).

The same argument works for $g_i$.

\section{QFI matrix for Gaussian states} \label{app:gaussian_qfi}
According to Williamson's theorem~\cite{Weedbrook2012Gaussian}, every covariance matrix can be diagonalised with a symplectic transformation: $V = S D S^\transp$, where $S \in Sp(2n)$ and $D = \mathrm{diag}(d_1,d_1,d_2,d_2,\dots,d_n,d_n)$ is the covariance matrix of a product of $n$ thermal states. A straightforward calculation gives the QFI matrix for a thermal state:
\begin{equation}
	V = D \Leftrightarrow F = \frac{1}{4}D^{-1}.
\end{equation}
From (\ref{eqn:qfi_formula}), we see that, under a linear transformation $V \to S V S^\transp$, the QFI matrix transforms in the same way: $F \to S F S^\transp$. Hence
\begin{equation}
	V = S D S^\transp \Leftrightarrow F = \frac{1}{4} S D^{-1} S^\transp.
\end{equation}
To arrive at the claimed expression, we use the fact that $S \Omega S^\transp = \Omega$, or $S \Omega = \Omega (S^\transp)^{-1}$:
\begin{align}
	\Omega V^{-1} \Omega^\transp & = \Omega (S^\transp)^{-1} D^{-1} S^{-1} \Omega^\transp \\
		& = S \Omega D^{-1} \Omega^\transp S^\transp \\
		& = S D^{-1} S \\
		& = 4F.
\end{align}

\section{Gaussian transformations} \label{app:gaussian}
A crucial property of Gaussian POVMs $E(\bg{\alpha}) = \pi^{-n} D(\bg{\alpha}) \Lambda D(\bg{\alpha})^\dagger$ is that every outcome has the same covariance matrix, independent of the measurement outcome $\bg{\alpha}$~\cite{Weedbrook2012Gaussian}. Therefore the same conditions apply to deterministic state transformations as to probabilistic ones.  This also  implies that it is always sufficient to consider a single measurement step.

\subsection{Single-mode $\mc{P}_0$ transformations}
\begin{figure}[h]
	\centering
	\includegraphics[scale=1]{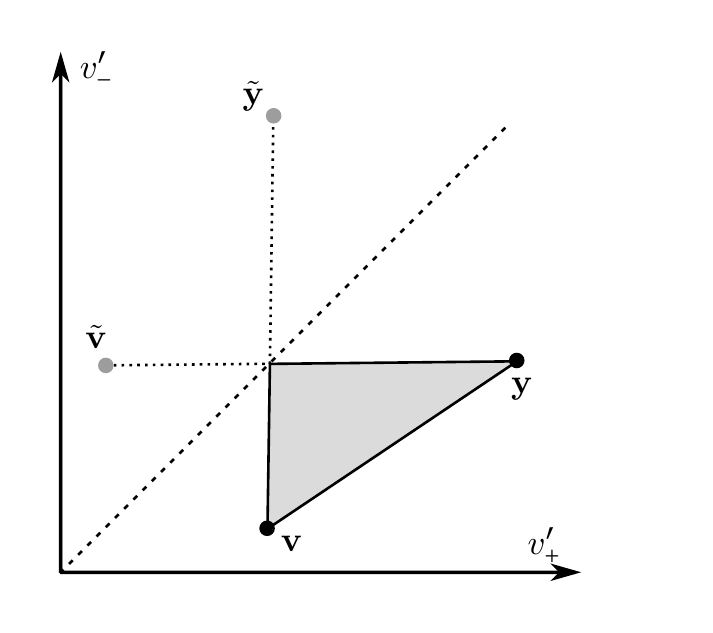}
	\caption{The region of achievable $\mathbf v'$ for fixed $\mathbf v$ and $\mathbf y$ is shaded. The points $\tilde{\mathbf v}$ and $\tilde{\mathbf y}$ are the reflections of $\mathbf v$ and $\mathbf y$ in the line $v'_+ = v'_-$.}
	\label{fig:triangle}
\end{figure}

We consider a single mode $S$ interacting with a single ancilla mode $A$ (which is sufficient by Lemma~\ref{lem:ancilla_size}), with covariance matrices $V$ and $Y$ respectively. From Lemma~\ref{lem:cs_decomp}, we can assume the PL unitary interaction to consist of a single beam splitter of reflectivity $\eta$, with phase rotations of each mode before and after. Thus the covariance matrix of the output can be written as
\begin{equation} \label{eqn:cov_sum}
	V' = (1-\eta) R_S V R_S^\transp + \eta R_A Y R_A^\transp, \quad R_S, R_A \in K(1).
\end{equation}
We will determine the set of $V'$ achievable by varying $\eta,R_S,R_A$. First consider fixing $\eta$. Given a pair of hermitian matrices $A,B$, there exists a set of linear inequalities constraining the eigenvalues of their sum $C=A+B$ in terms of the eigenvalues of $A$ and $B$ -- in the two-dimensional case, three inequalities are necessary and sufficient for the existence of the triple $(A,B,C)$~\cite{Bhatia1997Matrix} (Section III.2). Applied to (\ref{eqn:cov_sum}), these are
\begin{subequations}
	\begin{align}
		v_+' & \leq (1-\eta) v_+ + \eta y_+ \\
		v_-' & \leq (1-\eta) v_+ + \eta y_- \\
		v_-' & \leq (1-\eta) v_- + \eta y_+,
	\end{align}
\end{subequations}
where the eigenvalues of $V$ are $v_+ \geq v_-$ and similarly for $Y,V'$. To eliminate $\eta$, we use the fact that $\tr V' = (1-\eta) \tr V + \eta \tr Y$, i.e.,
\begin{equation}
	v'_+ + v'_- = (1-\eta)(v_+ + v_-) + \eta(y_+ + y_-).
\end{equation}
Thus we obtain the following necessary and sufficient conditions for the existence of $\eta,R_S,R_A$ satisfying (\ref{eqn:cov_sum}):
\begin{subequations}
	\begin{align}
		v'_+ (y_- - v_-) & \leq v'_- (y_+ - v_+) + v_+ y_- - v_- y_+ \\
		v'_- (y_+ - v_-) & \leq v'_+ (y_- - v_+) + v_+ y_+ - v_- y_- \\
		v'_- (y_- - v_+) & \leq v'_+ (y_+ - v_-) + v_- y_- - v_+ y_+.
	\end{align}
\end{subequations}
As illustrated in Fig.~\ref{fig:triangle}, this corresponds to a triangular region in the $(v'_+,v'_-)$ plane.

Now we use the ability to choose any classical ancilla, i.e., to choose any $y_+ \geq y_- \geq 1/2$. Clearly, every classical $V'$ is achievable by choosing any classical $Y$ and swapping the states using $\eta=1$. Given a nonclassical $V$ (meaning $v_- < 1/2$), the achievable nonclassical region is as shown in Fig.~\ref{fig:one_mode_gaussian}. This can be seen geometrically by considering triangles as in Fig.~\ref{fig:triangle}. It is bounded by two constraints: one situation where $y_+ \to \infty$, imposing that $v'_- \geq v_-$; the other where $y_+ = y_- = 1/2$. The inequality corresponding to the latter is found to be
\begin{equation}
	\frac{1/2-v'_-}{v'_+ - 1/2} \leq \frac{1/2-v_-}{v_+-1/2}.
\end{equation}
This determines the two necessary and sufficient monotones $\mc{N}_1,\mc{N}_2$ for $\mc{P}_0$ transformations.

\subsection{Single-mode $\gpnat$ transformations}
We first demonstrate the necessity of the monotones (assuming a nonclassical initial state). For $\mc{N}_1$, this follows immediately from Theorem~\ref{thm:qfi_gaussian} since $f_1 = (2 v_-)^{-1}$. For $\mc{N}_3$, we first convince ourselves that it suffices to consider a pure ancilla and measurement consisting of projection onto pure Gaussian states, i.e., with POVM elements of the form $\pi^{-1} D(\alpha) \proj{s} D(\alpha)^\dagger$, where $\ket{s}$ is a pure squeezed state. This works because any protocol with more general mixed elements can be obtained by a coarse-grained mixture of such pure cases.

We use the following result about Gaussian measurements at the level of covariance matrices~\cite{Eisert2002Distilling,Fiurasek2002Gaussian,Giedke2002Characterization}: suppose that a two-mode state has covariance matrix written in the $2 \times 2$ block form
\begin{equation} \label{eqn:block_cov}
	\begin{pmatrix}
		A & C \\
		C^\transp & B
	\end{pmatrix}.
\end{equation}
Then projection of the second mode onto a pure Gaussian state with covariance matrix $Z$ results in the first mode having the covariance matrix
\begin{equation} \label{eqn:schur_comp}
	A - C (B + Z)^{-1} C^\transp.
\end{equation}
In the present case, we start with system and ancilla covariance matrices $V_S,V_A$. We apply a beam splitter operation with reflectivity $\eta$, represented by
\begin{equation}
	R = \begin{pmatrix}
		\sqrt{1-\eta} I	& -\sqrt{\eta} I \\
		\sqrt{\eta} I	& \sqrt{1-\eta}I \\
	\end{pmatrix}.
\end{equation}
This results in $R(V_S \oplus V_A)R^\transp$, hence in (\ref{eqn:block_cov}) we have $A = (1-\eta) V_S + \eta V_A,\, B = \eta V_S + (1-\eta)V_A,\, C = \sqrt{\eta(1-\eta)} (V_S - V_A)$. The most general pure squeezed state has
\begin{equation}
	Z = \frac{1}{2} \begin{pmatrix}
		z^{-1} \cos^2 \theta + z \sin^2 \theta	&  (z^{-1} - z) \cos \theta \sin \theta \\
		 (z^{-1} - z) \cos \theta \sin \theta	& z^{-1} \sin^2 \theta + z \cos^2 \theta
	\end{pmatrix},
\end{equation}
where $z \geq 1$. Also note that, since we take $V_A = I/2$ (corresponding to a coherent state), $V_S$ and $V_A$ can be simultaneously diagonalised by single-mode phase rotations. Therefore, without loss of generality, we may write
\begin{align}
	A & = \begin{pmatrix}
		(1-\eta) v_+ + \eta/2 & 0 \\
		0 & (1-\eta)v_- + \eta/2
	\end{pmatrix}, \quad
	B = \begin{pmatrix}
	\eta v_+ + (1-\eta)/2 & 0 \\
	0 & \eta v_- + (1-\eta)/2
	\end{pmatrix}, \\
	C & = \begin{pmatrix}
		\sqrt{\eta(1-\eta)} (v_+ -1/2) & 0 \\
		0 & \sqrt{\eta(1-\eta)} (v_- -1/2)
	\end{pmatrix}.
\end{align}
We put these into (\ref{eqn:schur_comp}) and obtain (complicated) expressions for the eigenvalues $v'_+,v'_-$. Evaluating the expression for $\mc{N}_3$, and performing an optimisation over $\theta,\eta$ with all other parameters fixed, we obtain
\begin{equation}
	\max_{\theta,\eta} \frac{1/2-v'_-}{2-1/v'_+} = \frac{1/2-v_-}{2-1/v_+},
\end{equation}
attained at $\theta = 0, \eta=1/2$.\\

\begin{figure}[h]
	\centering
	\includegraphics[scale=1]{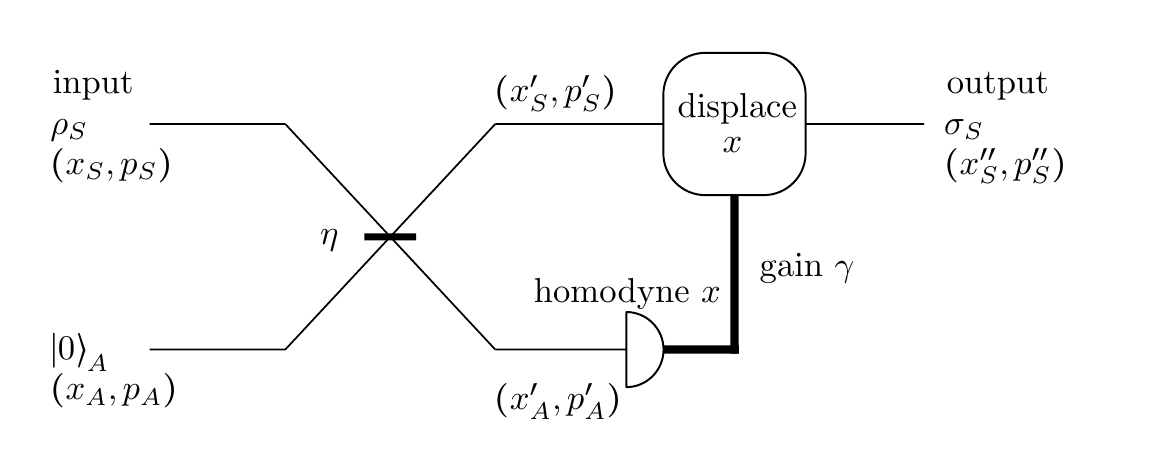}
	\caption{A $\gpnat$ operation that achieves a transformation with unchanged $\mc{N}_3$.}
	\label{fig:gaussian_pn}
\end{figure}

For a clearer physical picture, and to demonstrate sufficiency, we now give an explicit operation which takes $\rho \mapsto \sigma$ when $\mc{N}_3(\rho) = \mc{N}_3(\sigma)$. This also includes the conditional displacement on the output that is necessary to achieve a deterministic transformation. As illustrated in Fig.~\ref{fig:gaussian_pn}, this uses a vacuum ancilla, a beam splitter with reflectivity $\eta$ plus homodyne detection and feed-forward to a displacement with gain factor $\gamma$. The homodyne is performed on the most noisy quadrature.

We choose $x_S,p_S$ such that $V(\rho_S,x_S) = v_+, V(\rho_S,p_S) = v_-$ and perform a homodyne measurement of $x_A' = \sqrt{1-\eta} x_A - \sqrt{\eta} x_S$. Let the final displaced quadratures be $x_S'' = x_S' + \gamma x_A',\, p_S'' = p_S'$. We calculate the final variances as
\begin{equation} \label{eqn:vmin_final}
	v'_- = V(\sigma_S, p_S'') = (1-\eta) v_- + \eta/2.
\end{equation}
and $v'_+ = V(\sigma_S, x_S'')$. Using
\begin{align}
	x_S'' & = (\sqrt{1-\eta} x_S + \sqrt{\eta} x_A) + \gamma (-\sqrt{\eta} x_S + \sqrt{1-\eta} x_A) \\
		& = (\sqrt{1-\eta} - \gamma \sqrt{\eta}) x_S + (\sqrt{\eta} + \gamma \sqrt{1-\eta}) x_A
\end{align}
and noting that $x_S$ and $x_A$ are initially uncorrelated, we have
\begin{align}
	v'_+ = V(\sigma_S,x_S'') & = (\sqrt{1-\eta} - \gamma \sqrt{\eta})^2 V(\rho_S,x_S) + (\sqrt{\eta} + \gamma \sqrt{1-\eta})^2 V(\rho_A,x_A) \\
		& =  (\sqrt{1-\eta} - \gamma \sqrt{\eta})^2 v_+ + (\sqrt{\eta} + \gamma \sqrt{1-\eta})^2 \cdot \frac{1}{2}.
\end{align}
This expression is minimised by choosing a gain factor
\begin{equation}
	\gamma = \frac{\sqrt{\eta(1-\eta)}(v_+-1/2)}{\eta v_+ + (1-\eta)\cdot 1/2},
\end{equation}
from which one finds
\begin{equation}
	v'_+ = \frac{v_+ \cdot 1/2}{\eta v_+ + (1-\eta) \cdot 1/2} = \left( \frac{1-\eta}{v_+} + 2\eta \right)^{-1}.
\end{equation}
From this and (\ref{eqn:vmin_final}), we have
\begin{equation}
	\frac{2-1/v'_+}{2-1/v_+} = 1-\eta =\frac{1/2-v'_-}{1/2-v_-},
\end{equation}
showing that $\mc{N}_3$ is unchanged. Moreover, by choosing appropriate $\eta$, we see that $v'_+$ can be varied in the interval $[1/2,v_+]$. Hence any point lying on the curve $\mc{N}_3 = \text{constant}$ with $v'_- \geq v_-$ is reachable.

Finally, we can see that any point lying above the curves $\mc{N}_1,\mc{N}_3 = \text{constant}$ is reachable. After the above operation, one simply adds classical noise to $v'_-$ without affecting $v'_+$. In Fig.~\ref{fig:one_mode_gaussian}, this corresponds to moving vertically upwards in the plane.

\end{document}